%% file: main-asplos.tex
\documentclass[sigplan,screen]{acmart}
\AtBeginDocument{%
  }

\setcopyright{cc}
\setcctype{by}
\copyrightyear{2027}
\acmYear{2027}
\acmDOI{10.1145/3845814.3850079}
\acmConference[ASPLOS '27] {Proceedings of the 32nd ACM International Conference on Architectural Support for Programming Languages and Operating Systems, Volume 1}{April 11--15, 2027}{Heraklion, Greece.}
\acmBooktitle{Proceedings of the 32nd ACM International Conference on Architectural Support for Programming Languages and Operating Systems, Volume 1 (ASPLOS '27), April 11--15, 2027, Heraklion, Greece}
\acmISBN{979-8-4007-3012-2/2027/04}

\usepackage[labelformat=simple]{subcaption}
\usepackage{xspace}
\usepackage{algorithm}
\usepackage{algorithmic}
\usepackage{pifont}
\usepackage{balance}

\newcommand{\figref}[1]{Figure~\ref{#1}}
\newcommand{\work}{\textsc{NanoCP}\xspace}

\newif\ifrevhighlight
\revhighlightfalse   

\newcommand{\rev}[1]{%
  \ifrevhighlight
    \textcolor{blue}{#1}%
  \else
    #1%
  \fi
}

\newcommand{\blacknum}[1]{%
  \raisebox{-0.10em}{%
    \fontsize{9.5pt}{9.5pt}\selectfont
    \ding{\numexpr181+#1\relax}%
  }%
}

\begin{document}

\title[\work]{\work: Taming Load Imbalance in Data-Expert Parallel Decoding under Variable-Length Workloads}


\settopmatter{authorsperrow=3}

\author{Jiefei Chen}
\authornote{Equal contribution.}
\orcid{0009-0002-3832-6922}
\affiliation{
  \institution{Fudan University}
  \city{Shanghai}
  \country{China}
}
\affiliation{
  \institution{Shanghai AI Laboratory}
  \city{Shanghai}
  \country{China}
}

\author{Binbin Lin}
\authornotemark[1]
\orcid{0009-0007-3802-0399}
\affiliation{
  \institution{Shanghai Jiao Tong University}
  \city{Shanghai}
  \country{China}
}
\affiliation{
  \institution{Shanghai AI Laboratory}
  \city{Shanghai}
  \country{China}
}

\author{Jinming Ma}
\orcid{0009-0000-1351-1377}
\authornotemark[1]
\authornote{Corresponding authors.}
\affiliation{
  \institution{Shanghai AI Laboratory}
  \city{Shanghai}
  \country{China}
}
\email{jimyma@nano-infra.com}

\author{Jiangfei Duan}
\orcid{0000-0002-6327-2033}

\authornotemark[2]
\affiliation{
  \institution{The Chinese University of Hong Kong}
  \city{Hong Kong}
  \country{China}
}
\email{imjfduan@gmail.com}

\author{Haojie Duanmu}
\orcid{0009-0000-2118-3078}
\affiliation{
  \institution{Shanghai Jiao Tong University}
  \city{Shanghai}
  \country{China}
}
\affiliation{
  \institution{Shanghai AI Laboratory}
  \city{Shanghai}
  \country{China}
}

\author{Hao Liu}
\orcid{0009-0003-8799-3040}
\affiliation{
  \institution{Shanghai AI Laboratory}
  \city{Shanghai}
  \country{China}
}
\affiliation{
  \institution{Harbin Institute of Technology}
  \city{Harbin}
  \country{China}
}

\author{Qinxiu Cheng}
\orcid{0009-0000-4529-3888}
\affiliation{
  \institution{Shanghai AI Laboratory}
  \city{Shanghai}
  \country{China}
}

\author{Xiuhong Li}
\orcid{0000-0002-4896-121X}
\affiliation{
  \institution{Peking University}
  \city{Beijing}
  \country{China}
}

\author{Zhilin Pei}
\orcid{0009-0001-4663-205X}
\affiliation{
  \institution{Shanghai AI Laboratory}
  \city{Shanghai}
  \country{China}
}

\author{Hui Wang}
\orcid{0009-0002-0604-4621}
\authornotemark[2]
\affiliation{
  \institution{Shanghai AI Laboratory}
  \city{Shanghai}
  \country{China}
}
\email{wanghui@pjlab.org.cn}

\author{Xingcheng Zhang}
\orcid{0009-0006-8525-0608}
\affiliation{
  \institution{Shanghai AI Laboratory}
  \city{Shanghai}
  \country{China}
}
\affiliation{
  \institution{SenseTime}
  \city{Shanghai}
  \country{China}
}

\author{Dahua Lin}
\orcid{0000-0002-8865-7896}
\affiliation{
  \institution{Shanghai AI Laboratory}
  \city{Shanghai}
  \country{China}
}
\affiliation{
  \institution{The Chinese University of Hong Kong}
  \city{Hong Kong}
  \country{China}
}

\renewcommand{\shortauthors}{Jiefei Chen et al.}

\begin{abstract}

Modern serving systems for Mixture-of-Experts (MoE) models adopt hybrid data-expert parallelism: expert parallelism (EP) shards experts across GPUs to scale capacity, while data parallelism (DP) replicates attention layers across instances to process independent requests. Existing systems bind each request's attention, MoE communication, and KV cache to a single instance. Because attention latency scales with KV cache size while MoE communication latency scales with batch size, this binding cannot balance both simultaneously, producing EP stragglers; it also fragments KV memory across instances, inflating tail latency under long contexts. 
While existing context parallelism (CP) mitigates these constraints, existing group-level and uniform CP designs incur prohibitive communication and attention-side overheads.

We present \work, which decouples MoE communication from KV cache placement and achieves dual balance through dynamic context parallelism (DCP). DCP assigns each request a context-parallel degree sized to its KV footprint: long requests distribute attention across multiple instances; short requests remain local.

This dynamic parallelism effectively liquefies the KV cache
across DP instances within each node, balancing both the
per-instance KV cache occupancy and batch sizes without unnecessary
load-balancing costs.
To bridge DCP with static execution, \work introduces an ahead-of-time
(AOT) graph engine paired with a custom routing-based communication backend.
Experimental results show that, under 99\% TPOT-SLO
attainment, \work achieves 3.27$\times$ and 2.11$\times$
the maximum sustainable request rate of the best-performing
vLLM baseline on the DeepSeek-V3 Issue 1\% and Issue 5\%
workloads, respectively, while reducing P99 TPOT by
1.94$\times$ and 2.01$\times$  on the same two workloads, respectively.

\end{abstract}




\maketitle


\input{tex/intro-new-v2}
\input{tex/background}
\input{tex/system_overview}

\input{tex/control_plane}
\input{tex/data_plane}
\input{tex/evaluation}

\input{tex/discussion}
\input{tex/related}

\input{tex/conclusion}

\begin{acks}
We thank the anonymous reviewers and our shepherd for their feedback,
and Zihan Wang, Rongxin Cheng, and Qidong Su for insightful discussions
and suggestions.
This work was supported by the National Science and Technology Major
Project for New Generation Artificial Intelligence
(Project No. 2025ZD0123805).
\end{acks}

\bibliographystyle{ACM-Reference-Format}
\balance
\bibliography{bibfile}

\appendix

\include{tex/ae}









\end{document}

%% file: tex/intro-new-v2.tex
\section{Introduction}
\label{sec:intro}

Mixture-of-experts (MoE) architectures have become the dominant design for scaling large language models (LLMs) beyond the trillion-parameter regime without incurring proportional computation costs~\cite{dpskv3-arxiv-2024, kimiteam2026kimik25visualagentic, qwen3-arxiv-2024, jiang2024mixtral}. In production, these models serve requests of highly variable length, from short conversational turns to multi-million-token agentic and long-context workloads~\cite{gemini,deng2025swebenchpro,jimenez2023swebench}. Meeting tight token-level latency targets on such workloads is now a central system challenge in LLM inference.


Modern LLM serving systems typically adopt prefill-decode
disaggregation~\cite{zhong2024distserve, patel2024splitwise} together with
hybrid \textit{data-expert parallelism} (DP-EP)~\cite{insight,
vllm-sosp-2023, sglang-neurips-2024, dpskv3-arxiv-2024}.
In this design, non-expert layers (e.g., attention) are replicated across
data-parallel (DP) instances\footnote{A DP instance is the set of GPUs that
together hold one replica of the non-expert layers and one shard of the
experts.}, while experts are sharded across those instances through expert
parallelism (EP). Each MoE layer therefore incurs two all-to-all communication
phases, dispatch and combine, whose latency is determined by the slowest
participating rank under collective synchronization.
\rev{To describe request placement in this architecture, we introduce two
terms used throughout the paper. A request's \textit{MoE binding} denotes the
DP-instance set responsible for its MoE dispatch and combine operations. In
this paper, this set is a singleton. A request's \textit{KV binding} is the
set of one or more DP instances that store its KV cache and execute its
attention computation. In current request-level serving
systems~\cite{sglang-neurips-2024,vllm-sosp-2023,lumnix}, the MoE binding and
KV binding are tied to the same DP instance.}

\begin{figure*}[t]
    \centering
    \includegraphics[width=\linewidth]{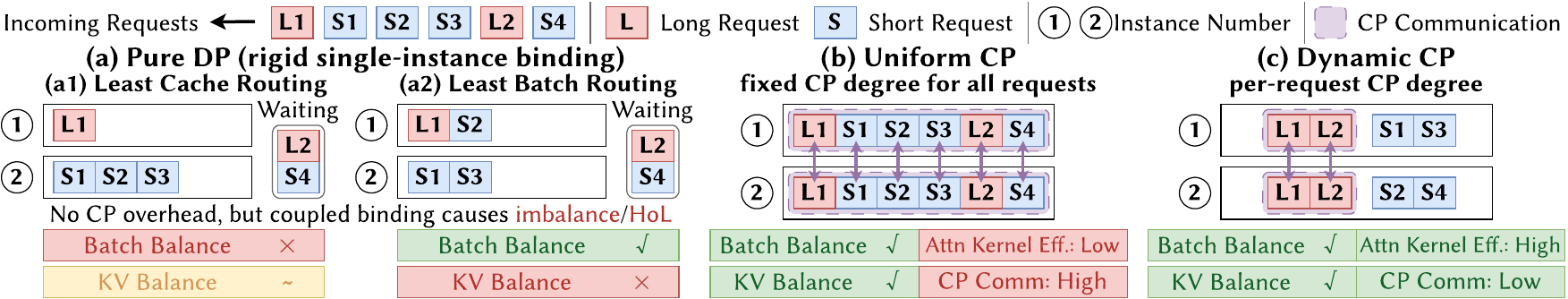}
    \Description{
Comparison of three request-placement strategies using long and short requests across two instances.
Pure data parallelism with LeastCache balances KV-cache usage but creates batch-size imbalance, while LeastBatch balances batch sizes but creates KV-cache imbalance.
Uniform context parallelism balances both dimensions but applies a fixed CP degree to all requests, causing low attention-kernel efficiency and high CP communication overhead.
Dynamic per-request CP assigns different CP degrees to requests and achieves both batch and KV-cache balance with high attention efficiency and low communication overhead.
}
    \caption{Existing load balancing strategies: (a1) least cache routes each incoming request to the instance with the largest number of free KV blocks, (a2) least batch routes it to the instance with the smallest running batch size, and 
    (b) \rev{uniform CP} achieves dual balance but incurs \rev{low attention-kernel efficiency} and high CP communication;
    (c) \work.}
    \label{fig:scheduling_comparison}
\end{figure*}

Under variable-length workloads, this design creates a fundamental scheduling problem. The two dominant phases in decoding scale differently: attention latency on each instance scales with its resident KV cache size~\cite{zhong2024distserve}, while dispatch and combine latency scale with the per-rank token batch entering the all-to-all~\cite{deepep2025}. As shown in Figure~\ref{fig:scheduling_comparison}(a), routing for KV cache balance inevitably skews batch sizes, and routing for batch size balance inevitably skews KV cache load. Under lock-step synchronization, both types of imbalance force faster ranks to wait for the slowest one, making end-to-end latency dominated by EP stragglers~\cite{chen2022ta, wang2025sp, wang2025harnessing, Oliaro2024FlexLLMTC,capacityawareinference}. This problem is orthogonal to prior work on expert load imbalance, such as EPLB and redundant experts~\cite{eplb}.

Tying the MoE binding and the KV binding to the same DP instance also creates a capacity-side problem. As MoE weights grow toward the trillion-parameter regime, model weights consume most of each instance's HBM, leaving KV cache capacity on the same order as a single long-context request. In one representative deployment---a 1T-parameter MoE on 32$\times$H200 with 32DP-32EP---the KV cache capacity of one instance is roughly 1M tokens, already comparable to the context windows of recent models~\cite{openaigpt54model, qwen3.5}. A single long request can therefore approach the capacity of an entire instance. As a result, even modest fragmentation can prevent the request from being scheduled, although the cluster still has enough aggregate free memory. This creates head-of-line (HoL) blocking and inflates tail latency, as shown in Figure~\ref{fig:scheduling_comparison}(a).


A related line of work is context parallelism (CP)~\cite{liu2024ringattention, infinitellm-arixv-2024, medha-arxiv-2025, ulysses, yang2025cp1m, jiang2025dcp, wu2025tokenlake, fgsp, wang2025flexsp, hotswitch, chen2025zeppelin, loongserve-sosp-2024, helix_parallel-arxiv-2025}, which expands a request's KV binding across multiple DP instances. However, existing CP systems are mainly designed for dense models or uniformly long-context serving, rather than variable-length MoE decoding.
\rev{LoongServe~\cite{loongserve-sosp-2024} elastically adjusts CP at the group level. Applied to DP-EP decoding, larger CP degrees for long-request groups reduce attention computation imbalance but cause batch size imbalance across DP instances, leading to EP stragglers.}
Helix~\cite{helix_parallel-arxiv-2025} instead applies the same CP degree to every request, as illustrated in Figure~\ref{fig:scheduling_comparison}(b).
\rev{It keeps one MoE binding per request but uniformly expands every request's KV binding to the same fixed-size group.}
This forces short requests to incur unnecessary cross-instance communication and increases the effective attention batch size on each CP instance, thereby reducing attention-kernel efficiency.
\rev{Neither LoongServe nor Helix independently selects each request's MoE binding and KV binding within the same decoding step, preventing them from jointly balancing attention computation and MoE communication with low CP overhead.}


To address this problem, we present \work, which introduces dynamic context parallelism (DCP) at per-request granularity to
\rev{decouple a request's MoE binding from its KV binding and schedule them independently.}
\rev{\work chooses each request's MoE binding to balance batch sizes across DP instances, while choosing each request's KV binding by trading off better attention computation balance against additional communication overhead.}
Under DCP, a request's CP degree is the number of DP instances in its KV binding. \work therefore assigns larger CP degrees to long requests to better balance attention computation, while assigning smaller CP degrees to short requests to avoid unnecessary communication. As illustrated in Figure~\ref{fig:scheduling_comparison}(c), this design achieves both batch size balance and KV cache balance 
while avoiding the attention-kernel efficiency loss and high CP communication overhead of uniform CP.

In the control plane, \work uses a centralized scheduler with a global view of load across instances, since local schedulers cannot jointly balance these dimensions or support token-level KV cache placement. Based on this global view, the scheduler determines, for each request, where MoE communication executes, what CP degree it uses, and how its KV cache is distributed across instances. However, executing this per-request flexibility creates highly asymmetric cross-instance traffic that conflicts with the static-shape assumptions of CUDA Graph capture and DeepEP's low-latency decode kernels~\cite{deepep2025}. To support this efficiently in the data plane, \work pairs an ahead-of-time (AOT) graph manager, which pre-compiles a bounded family of static-shape execution graphs, with a routing-based communication backend that issues asymmetric transfers outside the collective path.

In summary, this paper makes the following contributions:
\begin{enumerate}
\item We identify rigid request-to-instance binding as a fundamental source of both EP stragglers and memory fragmentation in hybrid DP-EP MoE serving, and show why existing CP designs are inefficient under variable-length workloads.
\item We propose \work, a \textit{dynamic context parallelism} (DCP) system that decouples each request's MoE binding from its KV binding within the same decoding step and supports per-request CP degrees under static-shape GPU runtimes.
\item We implement \work and show that under 99\%
TPOT-SLO attainment, it achieves 3.27$\times$ and
2.11$\times$ the maximum sustainable request rate of the
best-performing baseline on the DeepSeek-V3 Issue 1\% and
Issue 5\% workloads, respectively~\cite{vllm-sosp-2023,helix_parallel-arxiv-2025}.
At the request rate where the best-performing vLLM
baseline reaches 100\,ms mean TPOT, \work achieves
1.87$\times$--2.10$\times$ mean-TPOT speedups  on the
Issue 1\% and Issue 5\% workloads, evaluated separately
with DeepSeek-V3 and Kimi-K2.
\end{enumerate}

We have open-sourced \work. \footnote{\url{https://github.com/nano-infra/NanoEngine/tree/decentral-aug}}

%% file: tex/background.tex
\section{Background and Motivation}
\label{sec:back}

\subsection{Variable-Length Workloads}  
\label{sec:workload_analysis}  

Modern LLM serving systems must handle requests with highly variable sequence lengths, ranging from short conversational prompts~\cite{zhao2024wildchat, yan2025sharechat, zheng2023lmsyschat1m} to much longer multimodal and agentic workloads~\cite{gemini3pro2025, jimenez2023swebench, deng2025swebenchpro}. As shown by the OpenRouter statistics in Table~\ref{tab:length_distribution}~\cite{openrouter_rankings, aubakirova2026stateofai}, production traffic is highly skewed: most requests are short, but a non-negligible fraction (1.67\%) exceeds 100k tokens.
Following Medha~\cite{medha-arxiv-2025}, we construct evaluation traces by mixing ShareGPT-4o~\cite{cui2024sharegpt4o} for short requests with long-context software-engineering traces derived from GitHub issues for long requests. We use long-request ratios of 1\% and 5\% to represent both typical production workloads and heavier long-context stress cases.

\subsection{Data-Expert Parallel MoE Serving}
\label{sec:dp_serving}

\begin{figure}
    \centering
    \includegraphics[width=.9\columnwidth]{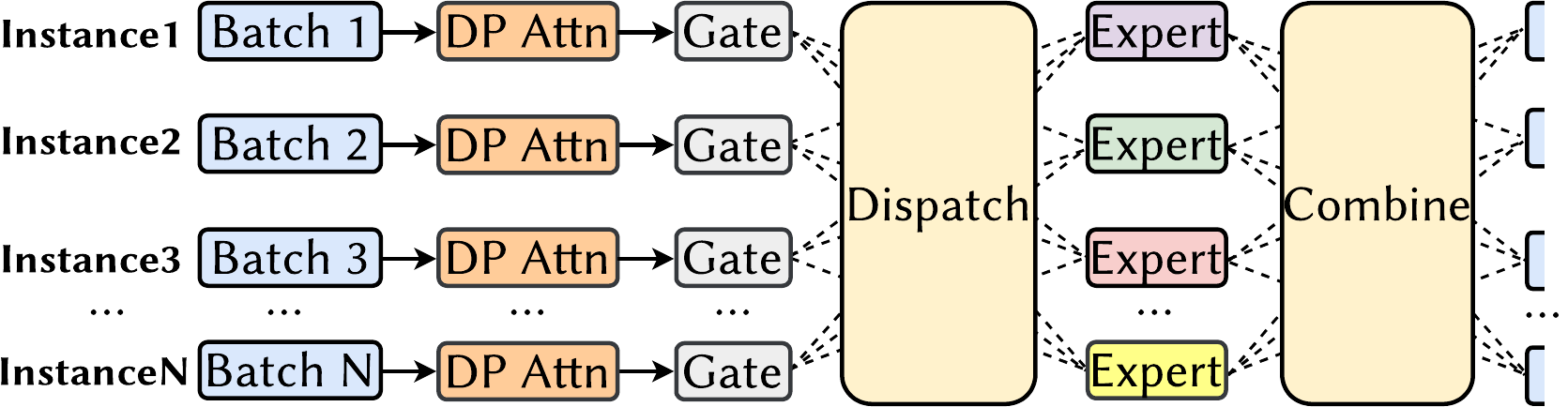}
    \caption{Data-expert parallel architecture.}
    \Description{
Data-expert parallel inference architecture with multiple data-parallel instances.
Each instance performs its replicated attention computation on a local request batch and applies a gating function.
Tokens are then exchanged through an all-to-all dispatch to sharded experts, followed by an all-to-all combine that returns expert outputs to the originating instances.
}
    \label{fig:dep_arch}
\end{figure}

As shown in Figure~\ref{fig:dep_arch}, modern MoE decoding
typically adopts data-expert parallelism (DP-EP) rather than
tensor parallelism (TP)~\cite{rajbhandari2022deepspeed,
megascale-infer,2025janus}. Head-wise TP is poorly suited to
Multi-Head Latent Attention (MLA)~\cite{dpskv3-arxiv-2024},
as it can replicate the KV cache and erode MLA's memory-efficiency
advantage~\cite{dpskv3-arxiv-2024,tpla}. Meanwhile, sharding
small experts produces fragmented GEMMs that fail to saturate
GPU compute units~\cite{megascale-infer}. DP-EP therefore
replicates attention layers across DP instances while sharding
experts through EP, enabling request-level parallelism and
leveraging aggregate memory bandwidth
~\cite{megascale-infer,deepep2025,2025janus}. Each MoE layer
performs one dispatch and one combine all-to-all.

\begin{figure}
    \centering
    \includegraphics[width=.9\linewidth]{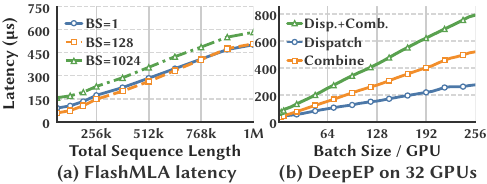}
    \caption{Microbenchmarks of (a) attention latency versus the total number of cached tokens per GPU under different batch sizes, where sequence length = total tokens / batch size and (b) DeepEP communication latency
    versus batch size.}
    \Description{
Two microbenchmark plots.
The left plot shows that FlashMLA attention latency increases with the total number of cached tokens per GPU. At matched cached-token counts, the largest tested batch size (BS=1024) incurs higher latency than BS=1 and BS=128.
The right plot shows that DeepEP dispatch and combine latency increase as the per-GPU batch size grows, with combine generally more expensive than dispatch.
}
    \label{fig:micro_attn_dpep}
\end{figure}


\paragraph{Different scaling factors in data-expert parallelism.}
A fundamental bottleneck in globally synchronized DP-EP
execution is \textit{instance-level imbalance}: attention
latency scales with the total KV cache size
(Figure~\ref{fig:micro_attn_dpep}(a)), whereas MoE communication
latency scales with batch size
(Figure~\ref{fig:micro_attn_dpep}(b)).
\rev{Figure~\ref{fig:micro_attn_dpep}(a) also shows that larger
effective batches can reduce attention-kernel efficiency at
the same KV-token count; we discuss this effect in
Section~\ref{sec:helix_parallel} when analyzing uniform CP.}

Under variable-length workloads, KV cache load and batch size
can diverge, making them difficult to balance simultaneously.
Lock-step synchronization then causes the slowest instance to
stall the entire DP-EP group. Prior work primarily addresses
\textit{expert-level imbalance} through expert placement
optimization~\cite{dpskv3-arxiv-2024,2025janus,
go2025moetuneroptimizedmixtureexpert,
li2026semanticparallel} or training-time load
balancing~\cite{switch-trans-JMLR-2022}. However, even perfectly
balanced experts do not eliminate this instance-level bottleneck.




\begin{table}
\centering
\caption{Request-length distributions (\%).}
\label{tab:length_distribution}

\small
\setlength{\tabcolsep}{4pt}
\begin{tabular}{@{}lccccc@{}}
\toprule
Dataset & < 1k & 1k--10k & 10k--100k & 100k--500k & 500k--1M \\
\midrule
OpenRouter\textsuperscript{1}
  & 31.82 & 50.08 & 16.42 & \multicolumn{2}{c}{1.67} \\
ShareGPT-4o
  & 85.7 & 10.7 & 3.5 & - & - \\
GitHub Issues
  & - & - & - & 65.06 & 34.94 \\
\bottomrule
\end{tabular}

\parbox{\columnwidth}{
\footnotesize
\textsuperscript{1} Sourced from
\url{https://openrouter.ai/rankings},
representing the call count records for Gemini 3 Flash Preview on Feb.~23.
}
\end{table}

\begin{figure}[t]
    \centering
    \includegraphics[width=1.0\linewidth]{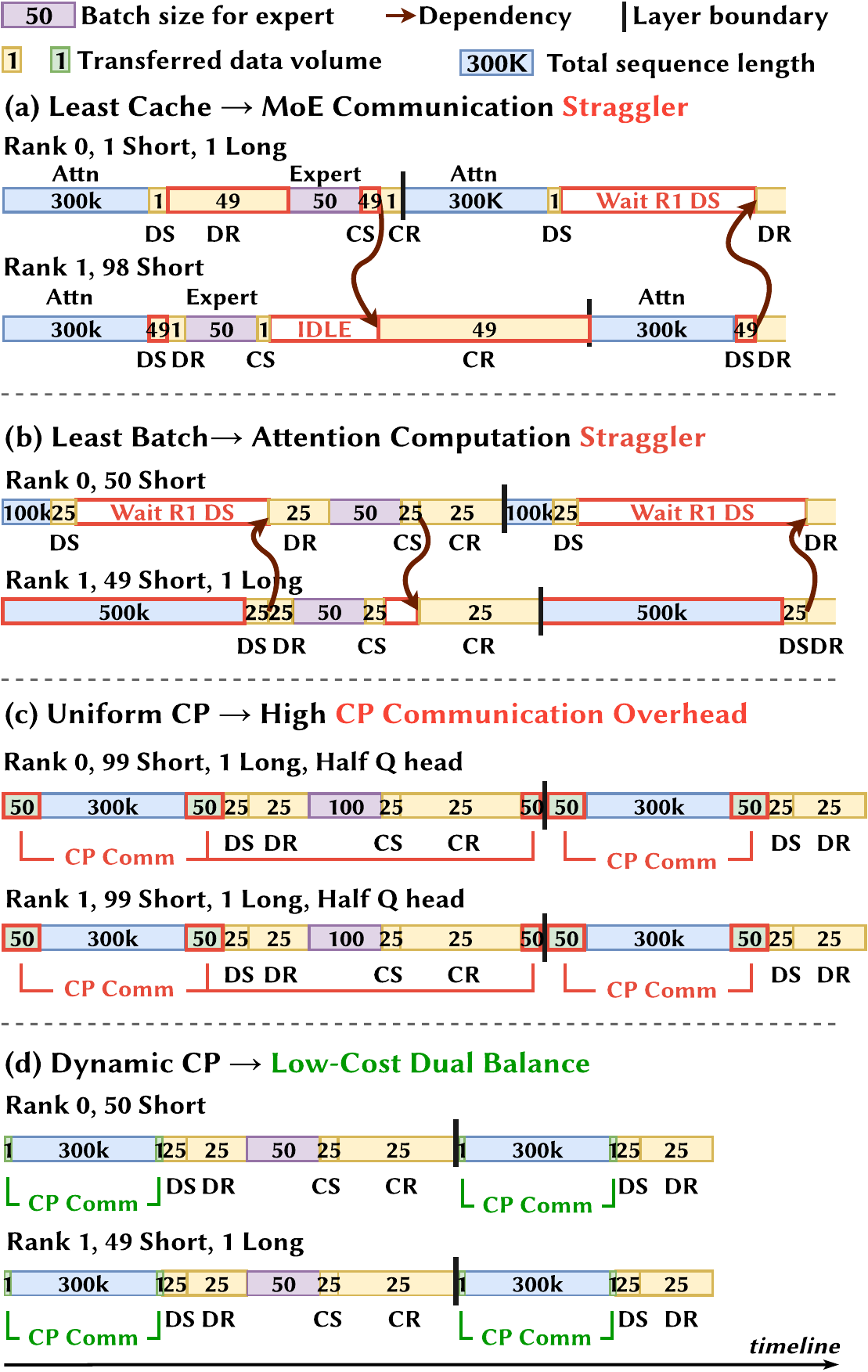}
    \caption{
    \rev{Per-layer timelines of four MoE decode strategies. (DS/DR: Dispatch Send/Receive; CS/CR: Combine Send/Receive).}}
    \Description{
Per-layer execution timelines for LeastCache, LeastBatch, uniform CP, and dynamic CP on two ranks.
LeastCache produces unequal MoE batch sizes and causes one rank to wait for the other's dispatch-receive completion.
LeastBatch produces unequal attention workloads and causes a rank with a shorter attention computation to wait for the slower rank.
Uniform CP adds CP communication before and after attention on both ranks.
Dynamic CP selectively applies low-cost CP communication and avoids the long waiting intervals of the other strategies.
}
    \label{fig:bubbles}
\end{figure}



\begin{figure*}
    \centering
    \includegraphics[width=.94\linewidth]{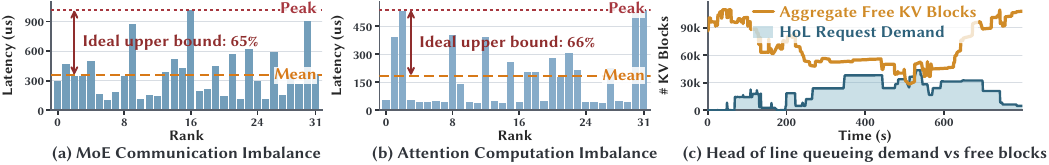}
    \caption{\rev{Limitations of existing request-level load balancing strategies.}}
    \Description{
Three measurements illustrating limitations of request-level load balancing.
The first plot shows large variation in MoE communication latency across 32 ranks, with one rank substantially slower than the mean.
The second shows a similarly large cross-rank imbalance in attention latency.
The third plots aggregate free KV-cache blocks and the memory demand of the head-of-line request over time, showing periods in which substantial aggregate free memory remains while a request is blocked because no single instance has sufficient local capacity.
}
    \label{fig:motivation_overview}
\end{figure*}


\subsection{Existing Load Balancing Strategies}
\label{sec:exist_balance_strategy}

\subsubsection{Request-Level Load Balancing}
\label{sec:req_level}

A common baseline for hybrid DP-EP serving is \emph{request-level load balancing}, which assigns each request to a single DP instance. As illustrated by the DP case in Figure~\ref{fig:binding_compare}(a), this design ties the request's MoE binding and KV binding to that same DP instance. This coupling is restrictive because attention scales with KV cache size, whereas MoE communication scales with batch size. As a result, this request-level scheduling can balance only one dimension.

\paragraph{LeastCache strategy and communication imbalance.}
This policy routes each incoming request to the DP instance with the smallest total KV cache size to balance attention computation. We implement this policy in vLLM for comparison. However, by ignoring batch size, it can create severe batch size imbalance, leading to communication imbalance and stragglers.
As illustrated in Figure~\ref{fig:bubbles}(a), even when attention loads are balanced, DP instance 0 may handle only 2 requests while DP instance 1 handles 98. DP instance 0 therefore spends much longer in dispatch receive~(DR), delaying its combine send~(CS). As a result, DP instance 1, despite finishing its MLP and combine send quickly, is blocked before combine receive~(CR) and remains idle. This straggler arises from batch size imbalance together with the sequential dependency between dispatch and combine.
Using 32 DeepSeek-V3 instances driven by the Issue 1\% trace in vLLM, with DP degree 32 and a constant arrival rate of 30 requests/s, Figure~\ref{fig:motivation_overview}(a) shows that MoE dispatch and combine latencies vary significantly across instances: most instances fall between 100 and 500~$\mu$s, while the straggler reaches 1020.6~$\mu$s. The mean is 354.7~$\mu$s, indicating that perfect load balance would reduce the maximum latency by 65.2\%.

\paragraph{LeastBatch strategy and attention computation imbalance.}
This policy routes each incoming request to the DP instance with the smallest batch size to balance MoE communication. vLLM's default scheduler follows this policy. However, because it ignores request-length heterogeneity, it can create severe KV cache imbalance and stragglers.
As shown in Figure~\ref{fig:bubbles}(b), even when batch sizes are balanced, attention loads can still differ substantially (e.g., 100K vs.\ 500K tokens). DP instance 0 finishes its lighter attention phase and dispatch send~(DS) quickly, but cannot proceed because DP instance 1 is still executing a heavier attention computation. This delays DP instance 1's dispatch send and leaves DP instance 0 idle. Thus, stragglers arise from KV cache imbalance together with the dependency between attention and dispatch.
Figure~\ref{fig:motivation_overview}(b) shows that attention latency also varies significantly under this policy: most instances remain around 40--55~$\mu$s, while the straggler reaches 539.5~$\mu$s. The mean is 184.3~$\mu$s, indicating that balancing attention computation would reduce the maximum latency by 65.8\%.

\paragraph{Head-of-line blocking.} Request-level routing also leads to severe head-of-line (HoL) blocking. Since each request must be placed on a single DP instance, a long-context request with a large KV cache cannot enter execution until one instance has enough local memory. Even when the cluster has sufficient aggregate free memory, the request can remain stuck at the head of the queue and block subsequent requests. Figure~\ref{fig:motivation_overview}(c) shows this effect: there is a clear gap between the high total free memory and the blocked request at the front of the queue. Because the KV cache of a request cannot be distributed across instances, aggregate free memory cannot be effectively utilized, leading to higher tail latency.

\subsubsection{Uniform CP}
\label{sec:helix_parallel}

\begin{figure}
    \centering
    \includegraphics[width=.9\linewidth]{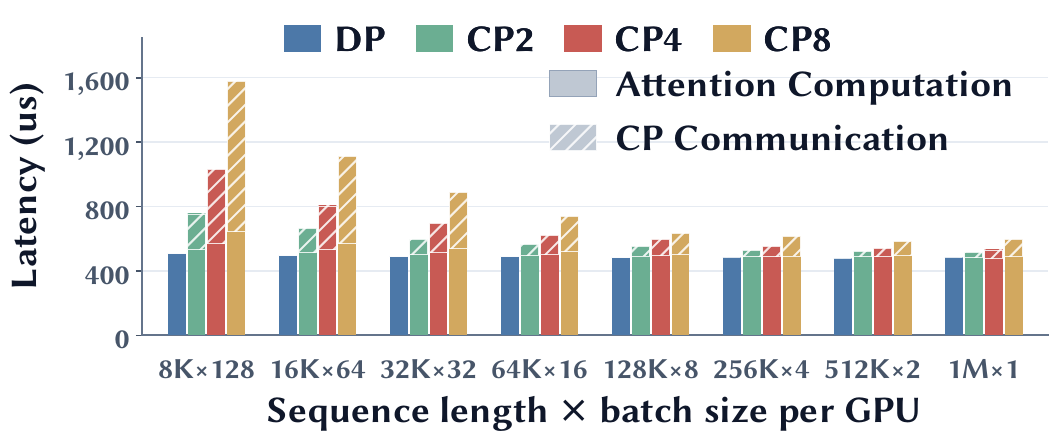}
    \caption{Per-layer attention latency breakdown of uniform CP
    under varying sequence lengths and batch sizes
    (total $\sim$1M tokens per GPU).}
    \Description{
Stacked-bar comparison of per-layer attention latency for data parallelism and uniform CP degrees of 2, 4, and 8 across sequence-length and batch-size combinations with roughly one million tokens per GPU.
Each bar separates attention computation from CP communication.
Uniform CP introduces additional communication at every CP degree, with the relative overhead especially large for shorter sequences and larger batches; the difference narrows as sequences become longer and batches smaller.
}
    \label{fig:vllm-dcp}
\end{figure}

\begin{figure}
    \centering
    \includegraphics[width=.95\linewidth]{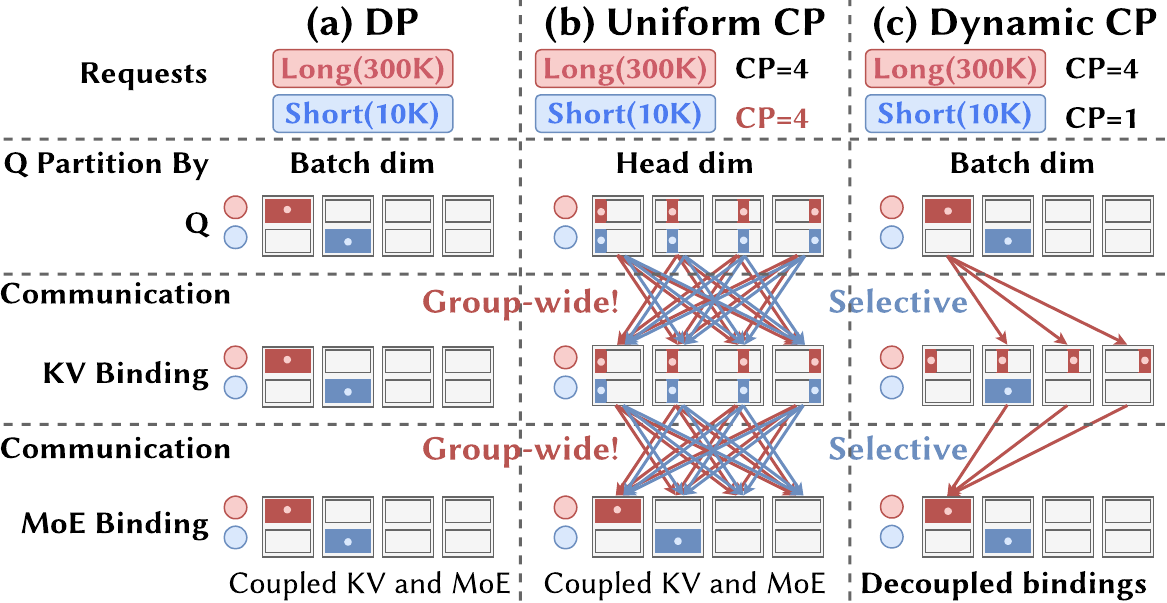}
    \caption{\rev{KV and MoE binding comparison.}}
    \Description{
Comparison of request bindings under data parallelism, uniform CP, and dynamic per-request CP for one long and one short request.
Data parallelism keeps each request's query, KV cache, and MoE execution on one instance.
Uniform CP assigns each request a singleton MoE binding and a KV binding spanning a fixed CP group, while applying the same CP degree to all requests.
Dynamic CP assigns the long request a larger KV binding and the short request a smaller one, uses selective communication, and decouples KV bindings from singleton MoE bindings.
}
    \label{fig:binding_compare}
\end{figure}

Helix~\cite{helix_parallel-arxiv-2025} is a representative
uniform CP design that improves over request-level scheduling
by allowing multiple GPUs to jointly execute each request's
attention, thereby distributing its KV cache and reducing
attention stragglers. As illustrated in
Figure~\ref{fig:binding_compare}(b), uniform CP keeps each
request's MoE binding on one DP instance while expanding its
KV binding to a fixed CP group.

However, as shown in Figure~\ref{fig:bubbles}(c), uniform CP
also forces short requests to participate in cross-instance
attention execution. Partitioning their KV cache and query
activations across the group requires routing queries,
returning partial attention results, and merging them into the
final output, introducing unnecessary communication overhead.

Figure~\ref{fig:vllm-dcp} quantifies this overhead under
different sequence-length and batch-size combinations.
Compared with DP, all CP configurations (CP2, CP4, and CP8)
incur substantially higher communication costs. The overhead
is especially pronounced for short sequences with large
batches (e.g., 8K$\times$128), where CP communication accounts
for a large fraction of the total attention latency.


\rev{Uniform CP can also reduce attention-kernel efficiency by
increasing the effective batch observed by each CP instance.
As shown in Figure~\ref{fig:micro_attn_dpep}(a), at a fixed
total KV-token count, a larger effective batch divides the
workload among more, shorter sequences. Because FlashMLA
uses a persistent tile scheduler, processing more sequences
incurs additional sequence-specific handling and tail-tile
inefficiency, whose costs are amortized over fewer KV tokens.
Here, BS denotes the effective attention-kernel batch after CP
redistribution, rather than the number of independent user
requests on one DP instance. \work avoids these
inefficiencies by keeping short requests local and assigning
larger CP degrees only to long requests.}

\subsection{DCP with Dual-Balanced Scheduling}
\label{sec:dcp_dual_balanced}

\rev{
Figure~\ref{fig:binding_compare} illustrates how the three designs differ in request binding. 
In DP, each request's KV binding and MoE binding are tied to the same instance.
Helix-style CP expands the KV binding of every request to a fixed CP group, so short and long requests both pay group-wide attention communication. 
\work keeps one MoE binding per request but chooses the KV binding size dynamically, allowing long requests to use CP while keeping short requests local.
}

\rev{
To jointly balance attention computation and MoE communication with low overhead, we propose dual-balanced scheduling with dynamic context parallelism (DCP). The key idea is to decouple a request’s MoE binding from its KV binding, so that MoE communication and KV cache placement can be scheduled independently within the same decoding step. As shown in Figure~\ref{fig:binding_compare}, \work selects each request’s singleton MoE binding to balance per-instance batch sizes for dispatch and combine, and independently selects its KV binding to balance KV cache and attention computation subject to cross-instance communication cost.
}

DCP enables this design efficiently in two ways. First, \work distributes the KV cache and attention computation of long requests across multiple instances only when doing so improves load balance, while keeping short requests local to avoid unnecessary communication. As illustrated in Figure~\ref{fig:bubbles}(d), \work makes Instance 0 and Instance 1 each hold 300K tokens of KV cache and process a batch size of 50, while incurring much lower communication overhead than Helix-style CP, leading to the shortest execution time in this example. Second, by partitioning a long request's KV cache within its selected KV binding, \work avoids requiring the request to fit into the free memory of a single instance, thereby utilizing fragmented free memory across instances and alleviating head-of-line blocking.

%% file: tex/system_overview.tex
\vspace{-.3em}
\section{\work Overview}
\label{sec:system_overview}

\begin{figure}[t]
    \centering
    \includegraphics[width=.9\linewidth]{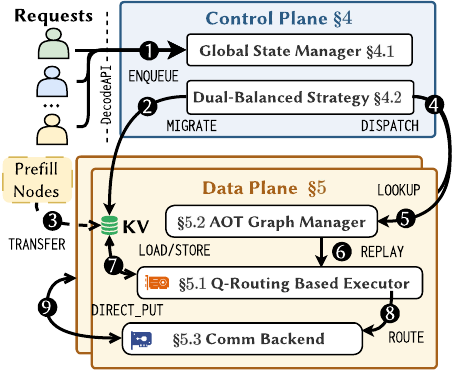}
    \caption{\work system overview.}
    \vspace{-1em}
    \Description{
System architecture divided into a control plane and a data plane.
Requests enter a global state manager, where a dual-balanced scheduler determines MoE and KV bindings and sends execution metadata to the data plane.
The data plane uses an ahead-of-time graph manager and a routing-based executor and communication backend.
During decoding, queries are routed from a request's MoE-binding instance to its KV-binding instances, which execute attention on local KV shards and directly return partial results for merging.
}
    \label{fig:overview}
\end{figure}

\figref{fig:overview} illustrates the \work{} architecture.
The control plane (Section~\ref{sec:control_plane}) maintains a \textit{global state manager} (Section~\ref{sec:global_state}) that pools all requests and GPU resources into a unified view.
Based on this view, the \textit{dual-balanced scheduling strategy} (Section~\ref{sec:dual-balanced}) generates per-iteration execution plans, determining each request's \textit{MoE binding} and \textit{KV binding}, and dispatches per-instance metadata to the data plane (Section~\ref{sec:data_plane}).
Since decoding relies on CUDA Graph to reduce CPU launch and synchronization overhead, per-request DCP's varying routing patterns conflict with its static-shape assumptions. \work{} therefore combines an \textit{AOT graph manager} (Section~\ref{sec:graph_convert}) with a \textit{routing-based communication backend} (Section~\ref{sec:pgas}) to support dynamic routing efficiently.

Specifically, a request's decode lifecycle flows through this architecture:
After completing external prefill, a request enters a global waiting queue (\blacknum{1}~\texttt{ENQUEUE}).
When admitted into decode, the scheduler determines its MoE binding, KV binding, and KV-token split.
Long requests may span multiple instances to distribute attention load and reduce KV cache imbalance, whereas short requests typically remain local to avoid unnecessary communication.
The control plane then triggers KV cache migration to the selected instances (\blacknum{2}~\texttt{MIGRATE}), while the data plane performs the transfer (\blacknum{3}~\texttt{TRANSFER}), and lowers the scheduling decision into per-instance routing metadata (\blacknum{4}~\texttt{DISPATCH}).
Each instance uses this metadata to look up (\blacknum{5}~\texttt{LOOKUP}) and replay (\blacknum{6}~\texttt{REPLAY}) the matching prebuilt graph.

During execution, the MoE binding sends the query tensor to the instances in the KV binding via the routing backend (\blacknum{8}~\texttt{ROUTE}).
Each instance performs attention over its local KV shard (\blacknum{7}~\texttt{LOAD/STORE}) and returns a partial result (\blacknum{9}~\texttt{DIRECT\_PUT}), which the MoE binding merges before proceeding to MoE execution.

%% file: tex/control_plane.tex
\section{\work Control Plane}
\label{sec:control_plane}


\subsection{Global State Management}
\label{sec:global_state}

Unlike mainstream LLM serving systems~\cite{vllm-sosp-2023,infinitellm-arixv-2024,lumnix,tropical}, which tightly couple the scheduler and executor within each data parallel (DP) instance, \work{} maintains a single global scheduler to coordinate all DP instances.
This design is driven by two requirements.
First, hybrid DP-EP execution requires strict \textit{per-iteration} load balancing, since the MoE dispatch and combine phases proceed in lock-step across DP instances and any imbalance immediately creates stragglers.
Second, DCP requires fine-grained resource allocation across instances, including token-level partitioning of a request's KV cache to reduce execution imbalance and memory fragmentation, which is difficult under coupled architectures where each scheduler manages only local resources.



\subsection{Dual-Balanced Scheduling}
\label{sec:dual-balanced}

\work realizes dual-balanced scheduling as a request-centric, length-aware scheduling algorithm. For each request, it determines two decoupled bindings, with one placing its MoE communication on a single instance and the other assigning its KV cache to a set of instances for attention computation. These decisions jointly balance MoE communication load and attention computation load across instances. Let \(\mathcal{N}\) denote the set of nodes and \(\mathcal{S}_n\) the CP instances on node \(n\).
For request \(r\), \(\ell_r\) is its sequence length, \(\mathcal{P}_r\) is the selected instance set for attention computation, \(m_r\) is the instance assigned to execute its MoE communication, and \(\textit{Split}_r\) maps each instance in \(\mathcal{P}_r\) to its assigned tokens.
For instance \(s\), \(K_s\) denotes its KV cache load, and $\mathbf{K}=(K_s)_{s}$ denotes the current KV-load state; \(B_s\) denotes the number of requests whose MoE communication is assigned to instance \(s\) in the current decoding iteration; \(B^{(n)}=\sum_{s\in\mathcal{S}_n}B_s\), and \(K_{\max}\) is the per-instance KV cache capacity. We use $Q_{act}$ to denote active requests with their existing
KV bindings $\mathcal{P}_r$, and $Q_{new}$ to denote requests
considered for admission.

\begin{algorithm}[t]
\caption{Dual-Balanced Scheduling Strategy}
\label{alg:dcp}
{\small
\begin{algorithmic}[1]
\REQUIRE $\mathcal{N}$, $Q_{new}$, $Q_{act}$, $\mathbf{K}$,
         $K_{\max}$, $\textsc{Bucket}(\cdot)$
\ENSURE Updated MoE assignments $m_r$ for active requests;
        $(\mathcal{P}_r, m_r, \textit{Split}_r)$ for each admitted
        $r\in Q_{new}$
        
 \textbf{Rebalance MoE Communication Assignments for Active Requests}
\STATE $B \leftarrow \mathbf{0}$
\FOR{each $r \in \textsc{SortByParticipantCountAsc}(Q_{act})$}
    \STATE $m_r \leftarrow \arg\min_{s \in \mathcal{P}_r} B_s$
    \STATE $B_{m_r} \leftarrow B_{m_r} + 1$
\ENDFOR

 \textbf{Node Selection and CP Degree}
\FOR{each $r \in Q_{new}$}
    \STATE $n^* \leftarrow \arg\min_{n \in \mathcal{N}} B^{(n)}$
    \STATE $k_r \leftarrow \min(\textsc{Bucket}(\ell_r),\, |\mathcal{S}_{n^*}|)$

 \textbf{Intra-Node Placement}
    \STATE $m_r \leftarrow \arg\min_{s \in \mathcal{S}_{n^*}} B_s$
    \STATE $\mathcal{C}_r \leftarrow \textsc{SelectSmallestKV}(\mathcal{S}_{n^*} \setminus \{m_r\},\, k_r - 1)$
    \STATE $\mathcal{P}_r \leftarrow \{m_r\} \cup \mathcal{C}_r$
    \STATE $\textit{Split}_r \leftarrow
        \textsc{WaterFill}(\mathcal{P}_r,\, \ell_r,\, \mathbf{K})$
    \IF{\textsc{CanAllocate}($\mathcal{P}_r$, $\textit{Split}_r$, $\mathbf{K}$, $K_{\max}$)}
        \STATE Commit $(n^*, \mathcal{P}_r, m_r, \textit{Split}_r)$ for request $r$
        \STATE Update local KV loads $K_s$ and MoE comm. load $B_{m_r}$
    \ELSE
        \STATE Keep $r$ in waiting queue for the next scheduling iteration
    \ENDIF
\ENDFOR
\end{algorithmic}
}
\end{algorithm}

\subsubsection{Rebalancing MoE Communication Assignments for Active Requests}
As requests complete, MoE communication assignments can become imbalanced. Since an active request can be reassigned within its existing KV binding \(\mathcal{P}_r\) without data migration, the scheduler processes requests in ascending order of \(|\mathcal{P}_r|\), prioritizing those with fewer reassignment choices, and assigns each request to the instance in \(\mathcal{P}_r\) with the fewest assigned requests (lines~2--3).

\subsubsection{Node Selection and CP Degree}
For each new request, the scheduler selects the node whose instances own the fewest total requests (line~7). It then determines the CP degree \(k_r\) using a length-bucket function \(\textsc{Bucket}(\ell_r)\) (line~8).
The key insight is that the optimal CP degree increases with request length: longer requests have larger KV cache footprints and thus lead to KV cache imbalance, so distributing them across more instances improves KV cache balance and reduces head-of-line blocking; shorter requests benefit less from CP and are more sensitive to its communication overhead.
Based on this insight, \(\textsc{Bucket}(\cdot)\) is derived from offline profiling. We sweep sequence lengths and candidate CP degrees, measure distributed attention latency under DCP, including both attention computation and communication, and select the CP degree that minimizes this latency for each length range. At runtime, the scheduler determines \(k_r\) by looking up \(\textsc{Bucket}(\ell_r)\). \rev{This mapping depends on the model and system configuration, rather than the request-length distribution.}

\subsubsection{Intra-Node Placement and KV Partitioning}
\rev{
Our implementation keeps each KV binding within one node to avoid critical-path cross-node traffic, while the abstraction supports cross-node placement. 
}

Within the selected node, the scheduler first chooses the instance that will execute the request's MoE communication (line~9). It then selects the remaining participating instances from those with the smallest KV cache load \(K_s\) (line~10), forming the final set \(\mathcal{P}_r\) for attention computation and KV placement (line~11).
Given this participant set, the scheduler computes a token split \(\textit{Split}_r\) using \(\textsc{WaterFill}\) (line~12), which distributes tokens to lower-loaded instances first. Specifically, \(\textsc{WaterFill}\) assigns tokens such that \(\sum_{s \in \mathcal{P}_r} \textit{Split}_r[s] = \ell_r\), while minimizing the peak post-allocation KV load \(\max_{s \in \mathcal{P}_r}(K_s + \textit{Split}_r[s])\).
The scheduler commits the placement only if \(\textsc{CanAllocate}\) confirms feasibility (line~13); otherwise, the request remains in the waiting queue for the next scheduling iteration (line~17).

\subsubsection{Routing Table Generation}
Once the scheduler finalizes placement decisions and token splits, it lowers them into per-instance $Q$-routing and Res/LSE routing tables, which decouple dynamic scheduling from data-plane execution. Section~\ref{sec:decoupled_data_path} details how the data plane consumes these tables to drive $Q$-routing execution (Figure~\ref{fig:block-table-transform}).

%% file: tex/data_plane.tex
\begin{figure}
    \centering
    \includegraphics[width=.85\linewidth]{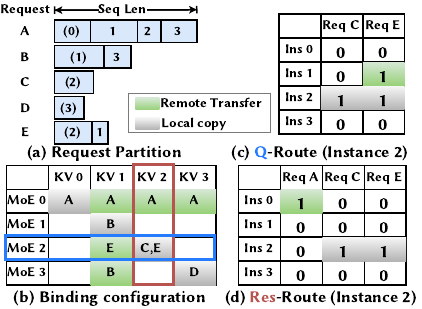}
    \caption{\rev{Routing table derivation. (a)~Request partition with MoE binding assignment (parenthesized). (b)~Binding configuration. (c)--(d)~Per-instance Q-Route and Res-Route tables for Instance~2.}}
    \Description{
Example showing how request placement is converted into routing tables for five requests across four instances.
The first panel shows each request's sequence partition and MoE-binding instance, and the second shows which instances hold each request's KV partitions.
For Instance 2, the Q-Route table marks the KV-binding destinations of locally generated queries, while the Res-Route table marks the MoE-binding destinations to which locally computed partial attention results are returned.
}
    \label{fig:block-table-transform}
\end{figure}


\section{\work Data Plane}
\label{sec:data_plane}

The DCP execution engine orchestrates the dual-balanced execution plan generated by the control plane.

\subsection{Decoupled Data-Path Design}
\label{sec:decoupled_data_path}

Given the execution plan generated by the control plane, \work uses a $Q$-routing data path between the \textit{MoE binding} and \textit{KV binding}: the singleton MoE binding performs MoE dispatch and combine, routes the lightweight query ($Q$) tensor to the KV-binding instances for distributed attention, and receives partial results for merging. This avoids migrating KV cache between the MoE and KV bindings during decoding. While similar $Q$-routing ideas have been explored in recent systems~\cite{loongserve-sosp-2024, infinitellm-arixv-2024, yang2025cp1m}, \work makes dynamic routing compatible with static GPU runtimes through an ahead-of-time (AOT) context-driven graph engine (Section~\ref{sec:graph_convert}) and a routing-based communication backend (Section~\ref{sec:pgas}).


Before each decoding iteration, the control plane derives per-instance routing tables from the KV cache distribution. 
For a given instance $s$, let $M$ and $N$ denote its local MoE and attention batch sizes, respectively: $M$ is the number of requests whose MoE binding includes $s$, and $N$ is the number whose KV binding includes $s$.
Figure~\ref{fig:block-table-transform} illustrates this conversion for five requests across four instances.
For example, C and E have Instance~2 as their MoE binding ($M{=}2$), while A, C, and E include Instance~2 in their KV bindings ($N{=}3$). Let $W_{cp}$ denote the number of instances in the DCP routing domain.  For each instance, the control plane derives two binary tables:
\emph{Q-Route} ($W_{cp} \times M$), indicating the KV-binding
instances to which each locally generated query is routed, and
\emph{Res-Route} ($W_{cp} \times N$), indicating the MoE-binding
instance to which each local partial result is returned.
Transfers are issued only for entries marked~$1$.

\begin{figure}
    \centering
    \includegraphics[width=.8\linewidth]{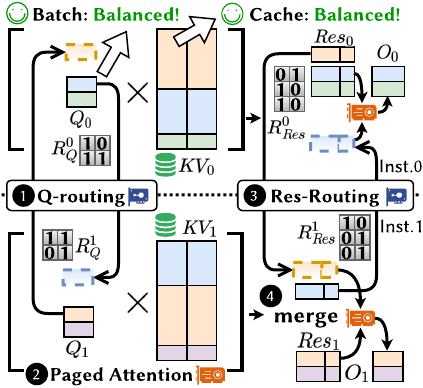}
    \caption{Execution process of \work DCP.}
    \Description{
Four-step execution workflow of dynamic context parallelism.
First, the MoE-binding instance projects queries and routes them to the selected KV-binding instances.
Second, each KV-binding instance performs paged attention using its local KV shard.
Third, partial attention outputs and normalization information are routed back to the MoE-binding instance.
Finally, the partial results are merged into the request's complete attention output.
The workflow simultaneously preserves balanced request batches and balanced KV-cache placement.
}
    \label{fig:workflow}
\end{figure}

Figure~\ref{fig:workflow} illustrates the execution workflow.
In \textbf{Phase 1} (\textit{Projection \& $Q$-Routing}), the MoE binding computes $Q$ and routes it to the KV binding via the Q-Route table (\blacknum{1}).
Each KV-binding instance then computes a partial attention output ($P$) and its log-sum-exp ($LSE$) scaling factor over its local KV partition (\textbf{Phase 2}, \blacknum{2}).
The partial results are returned via the Res-Route table (\textbf{Phase 3}, \blacknum{3}) and merged at the MoE binding using LSE merging~\cite{dao2023flashdecoding} (\textbf{Phase 4}, \blacknum{4}) before proceeding to the MoE layer.

\begin{algorithm}[t]
\caption{\rev{AOT Graph Capture and Replay}}
\label{alg:graph_dispatch}
{\small
\begin{algorithmic}[1]
\REQUIRE Execution-shape buckets $\mathcal{S}$ covering all supported $(M,N)$

\STATE \textbf{--- Initialization: Graph Capture ---}
\STATE $\mathcal{B} \leftarrow \textsc{AllocateSharedPool}(\mathcal{S})$
\FOR{each bucket $b=(\hat M,\hat N) \in \mathcal{S}$}
    \STATE $\mathcal{C}[b] \leftarrow \textsc{CreateView}(\mathcal{B}, b)$
    \STATE $\mathcal{G}[b] \leftarrow \textsc{CaptureGraph}(\mathcal{C}[b])$
\ENDFOR

\STATE \textbf{--- Runtime: Graph Replay ---}
\STATE $b \leftarrow \textsc{SmallestFit}((M,N), \mathcal{S})$
\STATE $\textsc{AsyncUpdate}(\mathcal{C}[b], \text{inputs and routing metadata})$
\STATE $\textsc{Replay}(\mathcal{G}[b])$

\end{algorithmic}
}
\end{algorithm}

\subsection{Bridging Dynamic Routing and Static Graphs}
\label{sec:graph_convert}

A fundamental challenge in implementing DCP is that per-request CP degrees
and KV bindings vary across iterations, changing both Q and Res routing and the
set of requests executing attention on each instance. Thus, the per-instance
execution shape $(M,N)$ is dynamic. A CUDA Graph, however, replays a
predetermined set of GPU operations with fixed launch shapes and buffer layouts.
CPU-side orchestration would introduce significant
synchronization overhead on the critical path~\cite{pytorch2023gptfast,ghosh2025pygraphrobustcompilersupport,Grape2023micro,KTransformers,Mustard}.
Alternatively, for each candidate $M$, capturing a graph that reserves the
worst-case attention batch size $N=M\cdot W_{cp}$ requires padding smaller
executions, inflating the effective attention batch and degrading
attention-kernel efficiency, as shown in
Section~\ref{sec:helix_parallel}.

\rev{
\work addresses this mismatch with an ahead-of-time (AOT) graph engine
(Algorithm~\ref{alg:graph_dispatch}). We define a bounded set $\mathcal{S}$
of two-dimensional execution-shape buckets $b=(\hat M,\hat N)$, such that
each supported runtime shape $(M,N)$ can be mapped to a bucket with
$\hat M \ge M$ and $\hat N \ge N$. During initialization, the engine
pre-allocates shared buffers large enough for the largest bucket. For each
bucket, it creates tensor views of the required sizes with fixed memory
addresses from these shared buffers and captures a complete DCP execution
graph over these views. At runtime, the engine selects the smallest bucket
that accommodates the current $(M,N)$, asynchronously updates its metadata,
and replays the corresponding graph. This design allows all bucket-specific
graphs to reuse the same underlying buffers
while avoiding per-operator CPU orchestration and limiting padding to the size of the selected bucket.
}


\subsection{Routing-Based Communication Backend}
\label{sec:pgas}

\begin{figure}
    \centering
    \includegraphics[width=.85\linewidth]{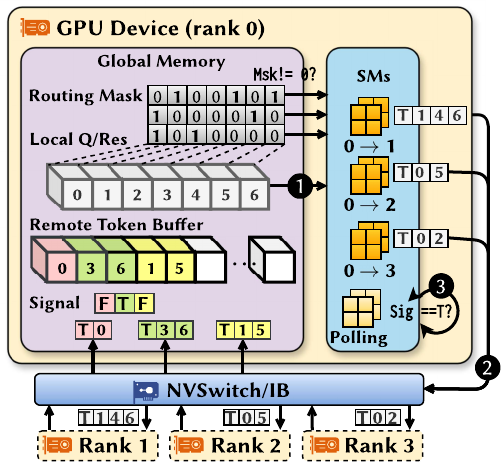}
    \caption{The microarchitecture of the routing-based communication backend.}
    \Description{
Microarchitecture of the routing-based communication backend on a GPU.
Routing masks, local query and result buffers, remote-token buffers, and completion signals reside in global memory.
GPU streaming multiprocessors read the routing mask to determine destination ranks and directly transfer selected data through NVSwitch or InfiniBand into preallocated buffers on remote GPUs.
Arrival signals allow receivers to detect completion without a dense collective communication operation.
}
    \label{fig:dlslime_design}
\end{figure}

To efficiently execute  dynamic routing between decoupled KV binding
and MoE binding within CUDA graphs, existing communication libraries do not match DCP’s communication pattern for two reasons. First, general collective
libraries~\cite{nccl_github,MSCCLang,msccl++,mooncake} are optimized for
dense exchanges with fixed layouts. Supporting DCP's sparse and asymmetric
transfers therefore requires excessive zero-padding, wasting bandwidth and
increasing latency. Second, specialized MoE
libraries~\cite{deepep2025,uccl-ep-osdi-2026,nccl-ep-arxiv-2026,pplx-kernels-arxiv-2025,eaas-arxiv-2025,stepfun2025step3largeaffordablemodelsystem}
are designed around fixed-width Top-$K$ token-to-expert routing. 
\rev{
In contrast,
DCP allows requests in the same batch to use different CP degrees, creating
variable per-request fan-out/fan-in. 
Mapping this pattern to fixed-width
Top-$K$ requires padding to the maximum CP degree or multiple communication invocations.
}
We therefore design a routing-based backend
tailored to DCP's sparse and variable-width routing.

Unlike traditional MoE routing, whose destinations depend on runtime gating outputs, DCP's communication topology is determined by the control plane before each iteration. The backend therefore uses the routing table to directly steer data transfers, avoiding translation into a fixed-width Top-$K$ representation and sending each data chunk directly to its target rank.

As illustrated in \figref{fig:dlslime_design}, the runtime 
interprets the routing table as a sparse $2$D mask matrix.
For each data chunk, the sender reads the corresponding mask entries from global memory (\blacknum{1}) to identify its target ranks, then transfers the payload directly into their preallocated receive buffers via NVLink for intra-node routing or InfiniBand (IB) for inter-node routing (\blacknum{2}). In parallel, the receiver polls a local arrival flag (\blacknum{3}).

%% file: tex/evaluation.tex
\begin{figure*}[t]
    \centering
    \includegraphics[width=1\linewidth]{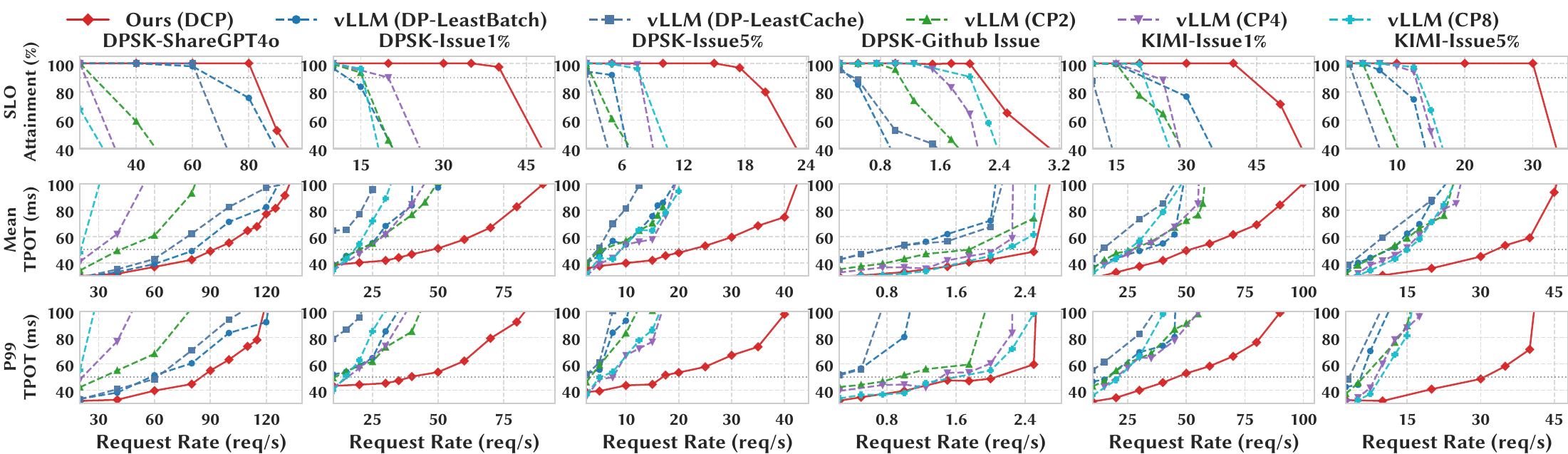}
    \caption{End-to-end generation performance on the H200 cluster.}
    \Description{
End-to-end performance across six workloads for the proposed DCP system and several vLLM data-parallel and uniform-CP configurations.
Each workload contains three plots showing SLO attainment, mean TPOT, and P99 TPOT as request rate increases.
The proposed system generally maintains high SLO attainment to higher request rates and lower mean and tail latency, with the largest advantages on workloads mixing short requests with one or five percent long-context requests.
Its performance is close to the strongest baselines on the pure short- and pure long-context workloads.
}
    \label{fig:metrics_comparison_deepseek}
\end{figure*}

\section{Evaluation}
\label{sec:eval}

\subsection{Experimental Setup}
\label{sec:eval-setup}

We implement \work{} in $\sim$10K lines of code atop FlashMLA~\cite{flashmla2025}, Triton-distributed~\cite{zheng2025tritondistributed}, DeepEP~\cite{deepep2025}, DeepGEMM~\cite{deepseek2025deepgemm}, nano-vllm~\cite{yu_nanovllm}, and Ray~\cite{ray-osdi-2018}. The control plane is written in C++ and sends scheduling metadata to the data plane via RDMA to minimize overhead. \work{} also supports multi-step execution~\cite{2023lmdeploy}, reusing one scheduling decision across multiple decoding iterations.

We evaluate \work on a cluster of NVIDIA H200 GPUs. Each node has eight GPUs connected by a fully-connected NVLink fabric (900\,GB/s bidirectional) and eight 50\,GB/s RDMA NICs for inter-node communication. We use DeepSeek-V3~\cite{dpskv3-arxiv-2024} and Kimi-K2~\cite{kimiteam2025kimik2openagentic} as model backbones.

\noindent\textbf{Workloads.}
We use ShareGPT-4o~\cite{cui2024sharegpt4o} as the short-context dataset and GitHub Issues~\cite{medha-arxiv-2025} as the long-context dataset. We synthesize mixed workloads by combining them at 1\% and 5\% long-context ratios, and also evaluate pure short- and long-context baselines.

\noindent\textbf{Baselines.}
Both \work and vLLM~\cite{vllm-sosp-2023} use equivalent kernels (FlashMLA for attention, DeepEP~\cite{deepep2025} for MoE all-to-all). All configurations use wide EP for expert layers. For attention, we compare:
\begin{itemize}
    \item \textbf{32DP:} Standard data-parallel attention across 32 GPUs with vLLM's default policy (LeastBatch) and an additional LeastCache policy.
    \item \textbf{DP-CP Hybrid:} 16DP2CP, 8DP4CP, and 4DP8CP, \rev{ using vLLM's native uniform CP with head-dimension partitioning, following Helix's uniform decode-CP data path~\cite{helix_parallel-arxiv-2025}. We do not include HOP-B described in Helix, as no directly usable open-source implementation is available.}
\end{itemize}

\noindent\textbf{Metrics.}
Our primary metric is time-per-output-token (TPOT) with a 50\,ms SLO target~\cite{DEEPSERVE,stepfun2025step3largeaffordablemodelsystem}. We report the maximum request rate at which each system sustains $\geq$99\% SLO attainment, i.e., at least 99\% of requests meet this TPOT target.


\begin{figure}
    \centering
    \includegraphics[width=\linewidth]{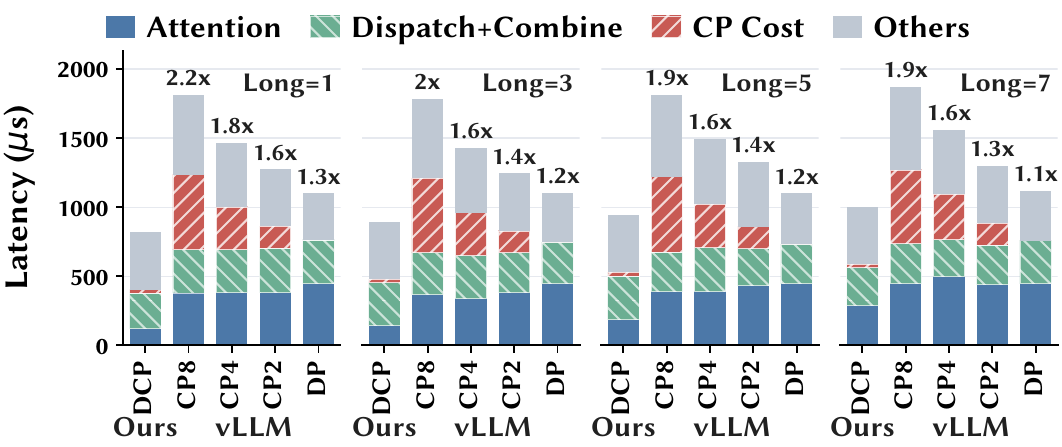}
    \caption{Per-layer latency breakdown with varying numbers of long requests per node.}
    \Description{
Stacked per-layer latency bars for the proposed DCP system and vLLM configurations using CP8, CP4, CP2, or pure data parallelism, evaluated with one, three, five, and seven long requests per node.
Latency is divided into attention, dispatch and combine, CP communication, and other costs.
The proposed system has the lowest total latency in all four cases by reducing attention imbalance while incurring substantially less CP communication than large uniform-CP configurations.
}
    \vspace{-.5em}
    \label{fig:microbenchmark}
\end{figure}

\begin{figure*}
    \centering
    \includegraphics[width=.9\linewidth]{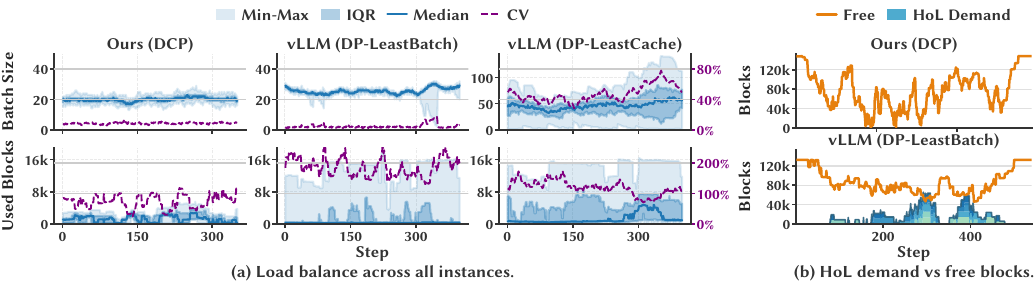}
    \caption{(a) Batch size and KV cache load balance across 32 instances under low load. (b) Head-of-line blocking under high load: free memory blocks vs.\ head-of-queue demand per node.}
    \Description{
Load-balance and head-of-line-blocking measurements.
Under low load, time-series plots compare batch size and KV-cache load across 32 instances for the proposed DCP system, vLLM LeastBatch, and vLLM LeastCache.
The proposed system keeps both batch-size and KV-cache variation relatively small, whereas LeastBatch has highly imbalanced KV-cache load and LeastCache has highly imbalanced batch sizes.
Under high load, the head-of-line plots show that vLLM can have blocked requests despite substantial aggregate free memory, while the proposed system mainly queues requests only when aggregate free memory is nearly exhausted.
}
    \label{fig:loadbalance}
\end{figure*}

\subsection{Micro Benchmark}
\label{sec:microbenchmark}

We compare \work against four vLLM baselines (vLLM 32DP, vLLM 16DP2CP, vLLM 8DP4CP, vLLM 4DP8CP) on 4 nodes, placing 64 short requests (2048 tokens) per GPU and 1, 3, 5, or 7 long requests (512K tokens) per node. For each setting, we report the latency breakdown of the slowest instance, since EP makes per-layer latency nearly identical across instances. Per-layer latency is decomposed into attention computation, MoE all-to-all, CP communication, and other costs.

Figure~\ref{fig:microbenchmark} shows that \work achieves the lowest overall latency across all four settings. Its CP communication overhead is much smaller than vLLM 4DP8CP's, because it enables CP only for requests that would otherwise create significant attention imbalance. \work also achieves the smallest maximum attention latency: compared with vLLM 32DP, it distributes long-request KV cache across instances rather than confining each to a single DP instance; compared with vLLM 4DP8CP, it avoids the inflated attention batch of uniform CP, where each CP instance executes attention over the full CP-group batch, hurting kernel efficiency. vLLM 16DP2CP and vLLM 8DP4CP suffer from both attention stragglers and reduced kernel efficiency.

The benefit of distributing long-request KV cache diminishes as the number of long requests increases per node: when they are few, this removes dominant attention stragglers; as their number grows, the room for further KV cache balancing shrinks.

\subsection{End-to-End Performance}
\label{sec:eval-end-to-end}

Figure~\ref{fig:metrics_comparison_deepseek} reports the end-to-end serving performance. Under mixed workloads, vLLM 32DP suffers from imbalances in both attention
computation and MoE dispatch/combine communication, leading to stragglers. 
 Uniform CP baselines partially alleviate this problem, but still underperform \work: larger CP groups better balance long requests but incur higher communication and enlarge the effective attention batch, reducing attention-kernel efficiency, whereas smaller CP groups reduce communication but leave substantial attention imbalance. 
By assigning CP degrees dynamically per request, \work mitigates these imbalances while avoiding the uniform overhead of fixed-degree CP. 
\rev{
On DeepSeek-V3, under 99\% SLO attainment, \work supports $3.27\times$ and $2.11\times$ the maximum sustainable request rate of the best-performing vLLM baseline on Issue 1\% and Issue 5\%, respectively.
}

Under the pure short-context workload (ShareGPT-4o), \work achieves slightly lower latency than vLLM 32DP, showing that selective DCP introduces negligible overhead when context parallelism is unnecessary.
Under the pure long-context workload (GitHub Issues), \work performs similarly to 4DP8CP. At such low request rates, CP8 adds little overhead and already reduces stragglers effectively.

\rev{
At the request rates where the best-performing baselines
reach a P99 TPOT of $100$\,ms, \work reduces P99 TPOT
by factors of $1.94\times$ and $2.01\times$ on Issue 1\%
and Issue 5\%, respectively. By distributing a request's KV cache across multiple DP instances, \work mitigates head-of-line blocking and prevents long-context requests from blocking subsequent ones.
}

\rev{
The gains also generalize to Kimi-K2. Under 99\% SLO
attainment, \work sustains $1.88\times$ and $2.20\times$
the maximum request rates of the best-performing vLLM
baselines on Issue 1\% and Issue 5\%, respectively.
At the request rates where the best-performing baselines
reach 100\,ms mean and P99 TPOT, respectively, \work
achieves $1.87\times$--$2.07\times$ mean-TPOT speedups
and $1.79\times$--$2.12\times$ P99-TPOT speedups.
These results show that DCP's
benefits are not specific to DeepSeek-V3.
}

%
\subsection{Load Balance Analysis}
\label{sec:eval-load-balance}

\work achieves better load balance and less HoL blocking than vLLM under both low and high load. We quantify cross-instance imbalance using the coefficient of variation (CV), defined as the standard deviation divided by the mean; lower CV indicates better load balance.

\noindent\textbf{Low load.}
\rev{
\figref{fig:loadbalance} shows that \work keeps batch-size CV close to LeastBatch (7.34\% vs. 3.24\%) while substantially reducing KV cache CV (51.15\% vs. 156.77\%). Compared with LeastCache, \work reduces both batch-size CV (7.34\% vs. 59.90\%) and KV cache CV (51.15\% vs. 87.30\%).
}

\noindent\textbf{High load.}
\figref{fig:loadbalance} compares \work against vLLM 32DP with its default scheduling policy. For each 8-GPU node, we compare the aggregate free memory blocks with the demand of the head-of-line request. In vLLM, substantial free memory often coexists with a blocked head-of-line request due to fragmentation. In \work, requests queue mainly when aggregate free memory is nearly exhausted, showing that distributing a request's KV cache across instances mitigates HoL blocking by utilizing fragmented memory.

\begin{figure}
    \centering
    \includegraphics[width=\linewidth]{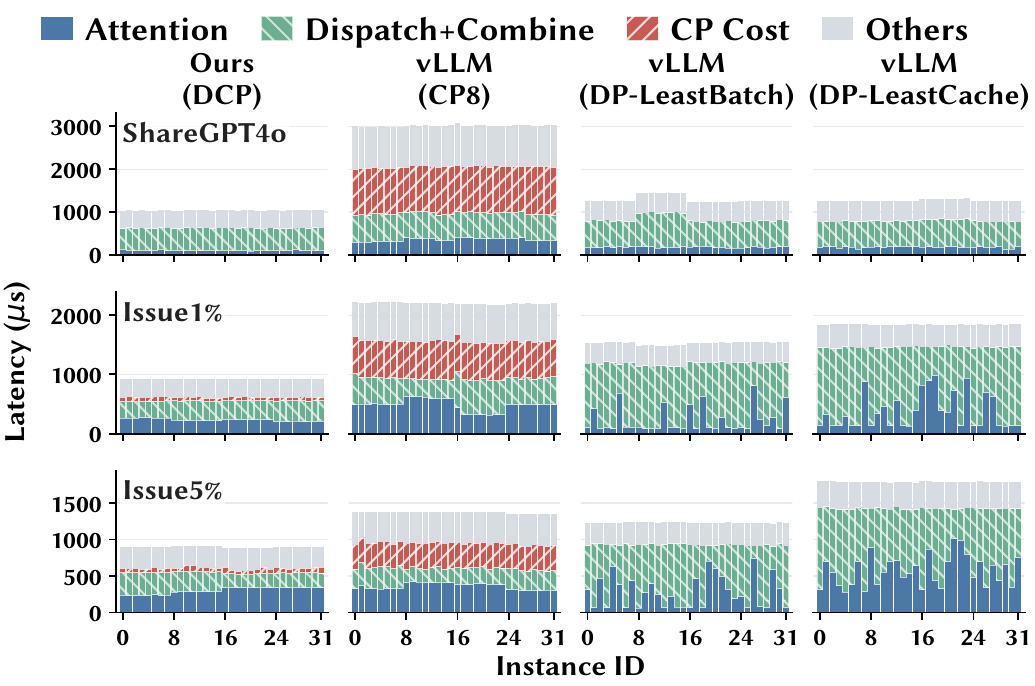}
    \caption{Latency breakdown of different routing and parallel strategies across 32 GPUs}
    \Description{
Per-instance latency breakdown across 32 GPUs for ShareGPT-4o and workloads containing one percent or five percent long-context requests.
Rows represent the three workloads and columns compare the proposed DCP system, uniform CP8, vLLM LeastBatch, and vLLM LeastCache.
Each instance's latency is divided into attention, dispatch and combine, CP communication, and other costs.
The proposed system shows relatively low and uniform latency across instances; CP8 incurs large CP communication costs, while the data-parallel baselines exhibit pronounced per-instance attention or MoE communication stragglers.
}
    \vspace{-.5em}
    
    \label{fig:microbench}
    
\end{figure}

\subsection{Latency Breakdown}

We decompose per-layer execution time into attention, MoE dispatch and combine, CP communication, and others across four configurations: \work, vLLM 4DP8CP, vLLM 32DP-LeastBatch, and vLLM 32DP-LeastCache. We evaluate ShareGPT-4o and its 1\%/5\% GitHub Issues mixtures by replaying requests extracted from serving logs across 32 instances (\figref{fig:microbench}).

Both vLLM 32DP baselines exhibit pronounced attention imbalance. Under ShareGPT-4o + 1\% GitHub Issues, vLLM 32DP LeastBatch exhibits a $7.26\times$ max-to-median latency gap (811.3$\mu$s vs. 111.8$\mu$s), while vLLM 32DP LeastCache still peaks at 986.065$\mu$s. By distributing long-context KV cache across instances, \work limits the maximum attention latency to 267.744$\mu$s and reduces the max-to-median gap by over 84\%.

MoE dispatch+combine shows a similar straggler pattern due to attention imbalance and batch size skew: vLLM 32DP LeastBatch and vLLM 32DP LeastCache reach maximum latencies of $1132.9\,\mu\text{s}$ and $1342.7\,\mu\text{s}$, respectively. By jointly balancing attention and MoE phases, \work lowers the maximum dispatch-and-combine latency to $350.9\,\mu\text{s}$.

By enabling DCP only for requests that need it, \work incurs much lower CP communication overhead than vLLM 4DP8CP: 60.4\,\(\mu\)s on average versus 629.8\,\(\mu\)s, a 90.41\% reduction.

\rev{
\subsection{Ablation Study}
}
\label{sec:ablation}

\begin{figure}[t]
    \centering
    \includegraphics[width=\linewidth]{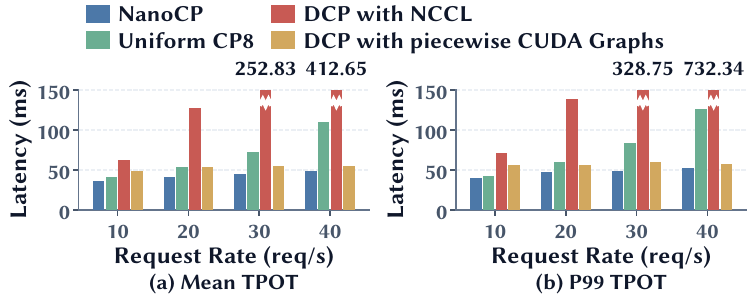}
    \caption{
        \rev{
        Ablation results on the DeepSeek-V3 Issue 1\%
        workload under different request rates.
        }
    }
    \Description{
Ablation study of mean and P99 TPOT at request rates from 10 to 40 requests per second.
The complete system is compared with uniform CP8, DCP using NCCL, and DCP using piecewise CUDA Graphs.
The complete system maintains the lowest latency as load increases.
Uniform CP8 degrades at high request rates, replacing the routing backend with NCCL causes a much larger latency increase, and piecewise CUDA Graphs remain close to but consistently slower than the complete system.
}
    \label{fig:tpot_ablation}
\end{figure}

We evaluate the contributions of per-request DCP and its
data-plane support on the DeepSeek-V3 Issue 1\% workload.
Figure~\ref{fig:tpot_ablation} reports mean and P99 TPOT
under different request rates.

\rev{
\noindent\textbf{Effectiveness of DCP.}
}
\rev{
As the request rate increases, \work maintains stable latency,
whereas uniform CP8 degrades rapidly. At 40 requests/s,
replacing DCP with uniform CP8 increases mean/P99 TPOT from
49.08/52.07\,ms to 109.97/126.77\,ms. This is because uniform
CP8 imposes unnecessary communication and reduces attention-kernel efficiency by inflating the effective attention batch on short requests.
}

\rev{
\noindent\textbf{Impact of data-plane support.}
}
\rev{
Replacing the routing backend with NCCL increases mean/P99
TPOT to 412.65/732.34\,ms at 40 requests/s, as dense
collectives poorly match DCP's sparse and asymmetric transfers.
Piecewise CUDA Graphs achieve 54.80/57.88\,ms, which are
11.7\%/11.2\% higher than \work, demonstrating the benefit
of capturing complete execution graphs specialized by the
two-dimensional execution shape $(M,N)$.
}

\subsection{System Overhead Analysis}
\label{sec:overhead}

\begin{figure}[t]
    \centering
    \includegraphics[width=.9\linewidth]{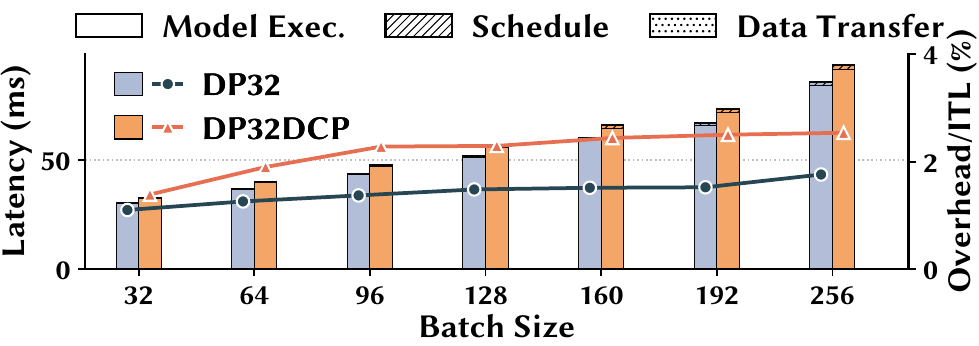}
    \caption{Schedule overhead comparison.}
    \Description{
Scheduler-overhead comparison for batch sizes from 32 to 256.
Stacked bars divide iteration latency into model execution, scheduling, and metadata transfer for standard DP32 and DP32 with dynamic CP.
Both total latency and scheduling overhead increase gradually with batch size, while the scheduling and transfer components remain a small fraction of total iteration time; dynamic CP adds only a modest overhead relative to DP32.
}
    \label{fig:schedule-overhead}
    \vspace{-1.em}
\end{figure}

\rev{
\noindent\textbf{Scheduler and AOT graph engine.}
}
We measure the centralized control-plane overhead on 32 GPUs with per-instance batch sizes from 32 to 256 (\figref{fig:schedule-overhead}). Under 32DP, scheduling plus RDMA metadata transfer accounts for only $1.17\%$--$1.77\%$ of iteration time. Enabling DCP raises this to $1.42\%$--$2.61\%$, an increase of at most $0.84$ percentage points.
\rev{
Graph selection is implemented as a simple dictionary lookup over prebuilt graphs indexed by execution-shape buckets, taking less than $2\mu\text{s}$ per iteration. After selecting the graph, \work performs one metadata injection per decode iteration. 
Under representative 4DP8CP settings from our end-to-end experiments, with per-GPU batch sizes up to 256 and Issue-1\% sequence lengths, metadata injection takes approximately $20\mu\text{s}$ per iteration.
}
The AOT graph engine prebuilds 48 CUDA Graphs consuming 5.32\,GiB per GPU (Table~\ref{tab:cuda_graph_memory}), an additional 4.30\,GiB ($\sim$3.0\% of H200 HBM) over vLLM 32DP, comparable to vLLM 4DP8CP (5.17\,GiB, 68 graphs).

\begin{figure}[t]
    \centering
    \includegraphics[width=.9\linewidth]{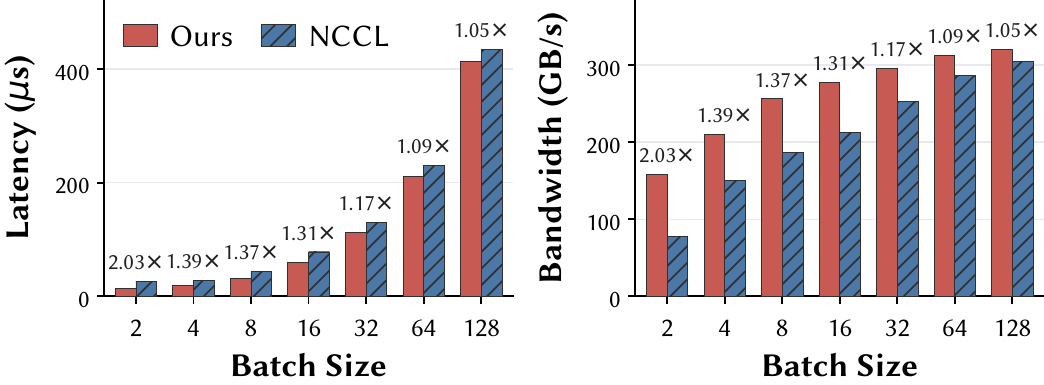}
    \caption{Performance of \work's routing-based backend vs.\ NCCL for query all-to-all transfer. Labels show speedup over NCCL.}
    \Description{
Two bar charts comparing the proposed routing-based communication backend with NCCL for query all-to-all transfers over batch sizes from 2 to 128.
The left chart shows lower communication latency for the proposed backend, and the right chart shows correspondingly higher effective bandwidth.
The advantage is largest at small batch sizes and gradually narrows as batch size increases, with labeled speedups ranging from about 2.03 times to 1.05 times.
}
    \label{fig:commlib}
\end{figure}

\noindent\textbf{Communication backend.}
Using DeepSeek-V3's query tensor (128 heads, head dimension 576, varying batch size), \work's routing-based backend accelerates query all-to-all
transfers by $1.05\times$--$2.03\times$ over NCCL across the
evaluated batch sizes (\figref{fig:commlib}).

\begin{figure}[t]
    \centering
    \includegraphics[width=.9\linewidth]{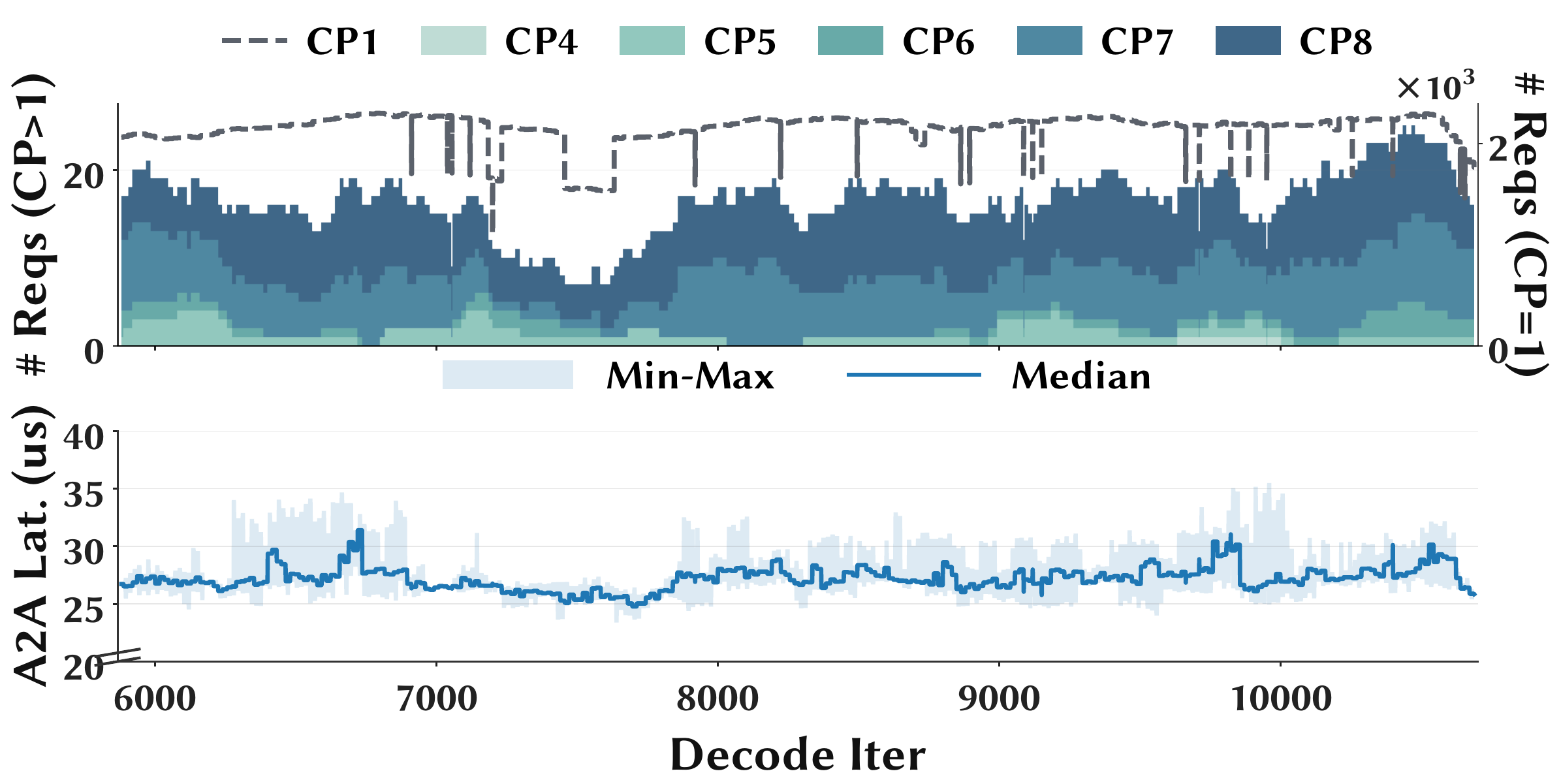}
    \caption{Runtime CP degrees and DCP communication latency.}
    \Description{
Runtime breakdown of dynamic CP usage over several thousand decode iterations.
The upper stacked area shows the small number of requests using cross-instance CP degrees from 4 through 8, while a dashed line on the secondary axis shows that the large majority of requests use CP degree 1.
The lower plot shows the median and min-to-max range of DCP all-to-all latency, which stays near a few tens of microseconds and peaks at roughly 35 microseconds.
}
    \label{fig:dcp_cost}
    \vspace{-1em}
\end{figure}

\noindent\textbf{DCP cost at runtime.}
As shown in \figref{fig:dcp_cost}, at peak only 25 of 2,279 requests (1.1\%) use cross-instance CP, with DCP all-to-all latency peaking at just 35.5\,$\mu$s.

\begin{table}[t]
\centering
\small
\setlength{\tabcolsep}{4pt}
\renewcommand{\arraystretch}{1.0}
\caption{CUDA Graph memory overhead comparison.}
\label{tab:cuda_graph_memory}
\begin{tabular}{@{\hspace{4pt}}lcc@{\hspace{4pt}}}
\toprule
Method & \# Graphs & Memory per GPU (GiB) \\
\midrule
vLLM (4DP8CP) & 68 & 5.17 \\
vLLM (32DP)     & 12 & 1.02 \\
\work            & 48 & 5.32 \\
\bottomrule
\end{tabular}
\end{table}


%% file: tex/discussion.tex
\section{Discussion}
\label{sec:discussion}

\begin{figure}[t]
    \centering
    \includegraphics[width=.9\linewidth]{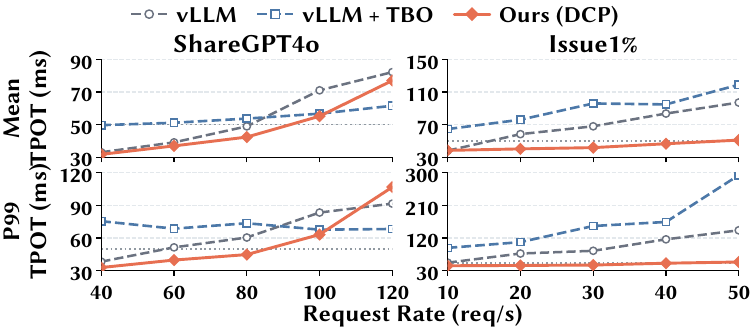}
    \caption{\rev{End-to-end performance with TBO.}}    
    \Description{
Mean and P99 TPOT versus request rate for vLLM, vLLM with two-batch overlap, and the proposed DCP system on ShareGPT-4o and the one-percent long-context workload.
On ShareGPT-4o, two-batch overlap improves vLLM latency, although the proposed DCP system remains competitive.
On the mixed long-context workload, two-batch overlap increases both mean and tail latency, while the proposed DCP system maintains substantially lower and more stable TPOT as request rate grows.
}
    \label{fig:vllm-tbo-e2e}
    \vspace{-1em}
\end{figure}

\noindent\textbf{Interaction with operation-level overlap.}


NanoFlow~\cite{nanoflow} and Two-Batch Overlap
(TBO)~\cite{deepseek_profile_data} overlap computation and communication
across sub-batches but leave request placement unchanged.
Variable-length workloads can therefore create instance-level stragglers
that shorten the overlap window. \work instead mitigates this imbalance
through dynamic KV cache placement and dual-balanced scheduling.

\rev{
The benefit of TBO depends on the workload. \figref{fig:vllm-tbo-e2e} shows that on ShareGPT-4o at
100 requests/s, it reduces mean TPOT from 82.33 to 56.70\,ms and
P99 TPOT from 93.61 to 67.67\,ms. On Issue 1\% at 10 requests/s,
however, it increases mean TPOT from 37.99 to 64.49\,ms and P99 TPOT
from 51.93 to 92.75\,ms, as long-context stragglers limit overlap
while microbatching introduces additional overhead. 
}

\rev{
TBO remains compatible with \work. Following DeepSeek's TBO pipeline
\cite{deepseek_profile_data}, we can insert Q-Route and Res/LSE-Route around
the attention stage, yielding
\emph{(SHARED + ATTN-0, Dispatch) $\rightarrow$ MLP $\rightarrow$
(Q-Route + ATTN-1 + Res/LSE-Route, Combine)}.
The two techniques optimize placement and overlap independently; joint
optimization is left to future work.
}


\rev{
\noindent\textbf{Generality and deployment scope.}
}
\rev{
Under DP-EP serving, MoE models with MLA, GQA, or MHA can all suffer
from imbalance in attention computation and MoE communication.
Increasing TP does not eliminate this imbalance. For GQA models with
few KV heads, scaling TP beyond the number of KV heads replicates KV
heads and their caches; for MLA, head-wise TP may replicate the shared
latent KV cache across TP ranks \cite{helix_parallel-arxiv-2025, mitra2025buzzpragmaticinferencedisaggregation}. Moreover, a moderate TP degree still
leaves multiple DP instances with replicated attention layers, whose
KV cache loads and batch sizes can diverge. SuperNode-class systems can
accommodate more such instances within a high-bandwidth domain, allowing
DCP to balance attention computation and MoE communication across more
instances while pooling more KV cache capacity.
}

\rev{
Our implementation confines each request's KV binding to one node.
On our H200 cluster, intra-node placement already achieves effective
load balance while avoiding Q/Res/LSE traffic on the inter-node critical
path. The scheduling abstraction and routing design also support
cross-node KV bindings, whose evaluation on SuperNode-class systems is
left to future work.
}



\rev{
\noindent\textbf{Scalability.}
}
\rev{
For larger deployments, \work can adopt a hierarchical design: a
lightweight global router assigns requests to GPU groups using aggregate
load summaries, while each group's scheduler performs fine-grained DCP
placement locally. We leave large-scale evaluation of this design to
future work.
To assess control-plane scalability, we use a scheduler-only CPU model to isolate the admission critical path across 4 and 32 logical nodes; each centralized/decentralized pair processes an identical request cohort. At four nodes, decentralized scheduling is comparable to centralized scheduling for Issue-1\%/R90 (5.5 versus 5.0\,ms) and is $1.36\times$ faster for Issue-5\%/R45 (18.0 versus 24.4\,ms). At 32 logical nodes, it reduces admission time from 119.5 to 13.6\,ms ($8.80\times$) and from 109.3 to 19.0\,ms ($5.76\times$), respectively. This widening gap indicates that sharding scheduling state across local schedulers mitigates the centralized admission bottleneck as the aggregate scheduling state grows.
}

%% file: tex/related.tex
\section{Related Work}
\label{sec:related}

\noindent\textbf{Context parallelism for LLM serving.}
Context parallelism (CP) was developed to scale sequence length in training~\cite{rajbhandari2022deepspeed,liu2024ringattention,chen2025zeppelin} and later adopted for serving. LoongServe~\cite{loongserve-sosp-2024}, InfiniteLLM~\cite{infinitellm-arixv-2024}, Meta's context-parallel inference design~\cite{yang2025cp1m}, Medha~\cite{medha-arxiv-2025}, and Helix~\cite{helix_parallel-arxiv-2025} use CP to pool distributed memory for long contexts or mitigate head-of-line blocking. 
\rev{
LoongServe elastically adjusts CP at the group level, while Helix applies a uniform CP degree across requests. Neither independently selects each request's MoE and KV bindings within the same decoding step, preventing joint balancing of attention load and MoE communication. 
}
\rev{
\work instead uses per-request dynamic CP to decouple KV cache placement from MoE execution, achieving dual balance while avoiding unnecessary CP overhead for short requests.
}

\noindent\textbf{MoE serving systems.}
Prior work optimizes MoE inference via efficient kernels and communication~\cite{gale2023megablocks,deepep2025,uccl-ep-osdi-2026,nccl-ep-arxiv-2026,pplx-kernels-arxiv-2025} or expert-level load balance~\cite{switch-trans-JMLR-2022,eaas-arxiv-2025,tang2025moentwine,2025janus}. MegaScale-Infer~\cite{megascale-infer} disaggregates attention and FFN but assumes balanced execution times. These optimizations are orthogonal to \work, which targets the DP-instance-level straggler effect under lock-step EP synchronization.

\noindent\textbf{Request scheduling in LLM serving.}
Existing systems~\cite{vllm-sosp-2023,sglang-neurips-2024,2023lmdeploy,tropical,orca,sarathi} bind each request's attention computation, MoE communication, and KV cache to a single DP instance. Llumnix~\cite{lumnix} adds cross-instance migration, while DistServe~\cite{zhong2024distserve}, Splitwise~\cite{patel2024splitwise}, and Mooncake~\cite{mooncake} disaggregate prefill and decode. 
\rev{
SBS~\cite{SBS} further performs load balancing for both prefill and decode, but its decode scheduler still assigns each request to a single DP instance based on batch size and KV cache load. 
}
These designs therefore retain a single request-to-instance binding and do not decouple a request's MoE binding from its KV binding within a decoding step. \work instead schedules the two bindings independently through dual-balanced scheduling with per-request CP degrees.

%% file: tex/conclusion.tex
\section{Conclusion}

\work addresses MoE inference conflicts by decoupling scheduling via dynamic KV cache placement and dual-balanced scheduling. Under the TPOT SLO, it improves maximum request rates by up to $3.27\times$ and significantly improves P99 TPOT over existing strategies.

%% file: tex/ae.tex
    \section{Artifact Appendix}
    
    \subsection{Abstract}
    The \work artifact includes the system implementation, benchmark scripts, workload traces
    and utilities, and plot generators. It supports the evaluation in this
    paper and provides recorded H200 latencies for reconstructing the
    CP-degree selection policy in Algorithm~\ref{alg:dcp}. Offline policy reconstruction
    requires only Python~3; serving experiments use the software and hardware
    configurations described below.
    
    \subsection{Artifact check-list (meta-information)}
    \begin{itemize}
      \setlength{\itemsep}{0pt}
      \setlength{\parskip}{0pt}
      \item \textbf{Methods:} \work (Python/C++), dynamic
            context parallelism, dual-balanced scheduling, offline policy
            search, and vLLM baselines.
      \item \textbf{Models:}
            \href{https://huggingface.co/deepseek-ai/DeepSeek-V3/tree/main}{DeepSeek-V3}
            and
            \href{https://huggingface.co/moonshotai/Kimi-K2-Instruct-0905}{Kimi-K2-Instruct-0905}.
      \item \textbf{Workloads:} ShareGPT-4o and GitHub Issues traces,
            including 1\% and 5\% long-context mixtures.
      \item \textbf{Runtime:} Linux; CUDA; PyTorch; Ray;
            NVLink and RDMA; modified vLLM; pinned DeepEP, DeepGEMM,
            FlashMLA, and DLSlime.
      \item \textbf{Hardware:} The full serving evaluation uses 32 GPUs across four nodes. See the figure-specific READMEs
            for component-test requirements.
      \item \textbf{Outputs:} mean/P99 time per output token (TPOT),
            50-ms TPOT-SLO attainment, request rate, component
            latencies, logs, and generated figures.
      \item \textbf{License:} MIT for released source; third-party
            models and data retain their upstream licenses.
    \end{itemize}
    
    \subsection{Access and dependencies}
    \textbf{Public archive.}
    The artifact is archived on
    \href{https://doi.org/10.5281/zenodo.23215463}{Zenodo}.
    Start with \path{ae_scripts/README.md}; each figure-specific README
    provides the commands, resource requirements, validation procedures,
    and expected outputs.
    
    \textbf{Models and environment.}
    Model checkpoints must be obtained separately. The artifact provides
    a Dockerfile specifying the pinned software environment, including
    the modified vLLM baseline. Model and dataset paths are configured
    in \texttt{paths.env}, using the example under \texttt{ae\_scripts/}.
    Distributed experiments use Linux with CUDA, GPU-aware RDMA, and
    identical absolute paths across workers. Storage requirements are
    dominated by model checkpoints and generated logs.
    
    \textbf{Data sets.}
    The artifact includes four preprocessed request-length traces under
    \texttt{ae\_scripts/dataset/}: ShareGPT-4o, GitHub Issues, and two mixed
    workloads containing 1\% and 5\% long-context requests, respectively.
    
    \subsection{Installation and basic testing}
    Fetch the archived source; run commands from \path{ae_scripts/}.

    \textbf{Evaluated environment.}
    The experiments reported in the paper were conducted on four nodes,
    each equipped with eight NVIDIA H200 GPUs with 141~GB of memory,
    intra-node NVLink, and inter-node RDMA. The evaluated software environment is specified by the Dockerfile under \path{ae_scripts/docker/}.
    
    \textbf{Deployment on other clusters.}
    Follow \path{ae_scripts/docker/README.md} to build the image and
    \path{ae_scripts/README.md} to create the containers and configure
    shared paths, the Ray cluster, worker SSH access, and RDMA settings.
    Copy \texttt{paths.env.example} to \texttt{paths.env} and configure
    the paths required by the selected experiments consistently across
    all nodes. Adjust the documented network addresses and RDMA settings
    to match the target cluster. The offline policy test below requires
    only Python~3 and runs without GPUs or the container.
    
    \textbf{Offline policy test (no GPUs required):}
    \begin{verbatim}
    python3 algorithm/search_sp_buckets.py
    \end{verbatim}
    Using the included grid of 258 sequence lengths and eight CP degrees,
    this script validates the data, reports the length-to-degree mapping,
    and saves the JSON policy to
    \path{algorithm/reproduced_policy.json}. GPU remeasurement is optional.
    
    \textbf{Distributed functional test (16 H200 GPUs):}
    After configuring two nodes, run:
    \begin{verbatim}
    python3 basic-test/run_basic_test.py
    \end{verbatim}
    Follow \path{basic-test/README.md} for the test procedure and expected
    results.
    
    \subsection{Experiments and expected results}
    Each figure README gives its measurement and plotting procedures.
    \textbf{Algorithm~\ref{alg:dcp}} uses \path{algorithm/} for offline policy
    reconstruction.
    \textbf{Figures~\ref{fig:micro_attn_dpep},
    \ref{fig:motivation_overview}, and \ref{fig:vllm-dcp}}
    use \path{fig3/}, \path{fig5/}, and \path{fig6/}
    for operator and imbalance measurements.
    \textbf{Figures~\ref{fig:metrics_comparison_deepseek}--%
    \ref{fig:schedule-overhead} and \ref{fig:vllm-tbo-e2e}}
    use the corresponding \path{fig*/} workflows for serving comparisons,
    profiling, and ablations on 32 GPUs;
    the complete Figure~\ref{fig:metrics_comparison_deepseek} sweep
    takes approximately four days.
    \textbf{Figures~\ref{fig:commlib}--\ref{fig:dcp_cost}}
    measure collective communication on eight GPUs,
    with Figure~\ref{fig:dcp_cost} additionally requiring
    a four-node trace capture.
    
    For DeepSeek-V3 at $\geq99\%$ attainment of the 50-ms TPOT SLO,
    \work sustains $3.27\times$ and $2.11\times$ the best evaluated vLLM
    request rates on Issue~1\% and Issue~5\%, respectively (Figure~\ref{fig:metrics_comparison_deepseek}).
    Results may vary with hardware, software, and load.